\documentclass{naturep}

\usepackage{amssymb}
\usepackage{amsmath}
\usepackage{graphicx}
\usepackage{algorithmicx}
\usepackage{algorithm}
\usepackage{subcaption}
\usepackage[noend]{algpseudocode}
\usepackage{bbm}
\usepackage{lineno}
\usepackage[margin=0.6in]{geometry}
\usepackage{ragged2e}
\usepackage{setspace}
\usepackage{longtable}
\usepackage{caption}
\usepackage{hyperref}
\usepackage{caption}
\usepackage{anyfontsize}
\usepackage{setspace}
\usepackage{float}
\usepackage{booktabs}
\usepackage{multirow}
\usepackage{xcolor}

\makeatletter
\let\saved@includegraphics\includegraphics
\AtBeginDocument{\let\includegraphics\saved@includegraphics}

\makeatother

\title{\begin{flushleft}{\begin{spacing}{1}Deep Learning-based Computational Pathology Predicts Origins for Cancers of Unknown Primary\end{spacing}}\end{flushleft}}

\begin{document}

\maketitle
\begin{spacing}{1.8}
\vspace{-15mm}
\noindent Ming Y. Lu$^{1,3,4}$, Melissa Zhao$^{1}$, Maha Shady$^{1,2,3}$, Jana Lipkova$^{1,3,4}$, Tiffany Y. Chen$^{1,3,5}$,\\ Drew F. K. Williamson$^{1,3,5}$, and Faisal Mahmood$^{*1,3,4}$
\begin{affiliations}
 \item Department of Pathology, Brigham and Women's Hospital, Harvard Medical School, Boston, MA
 \item Department of Biomedical Informatics, Harvard Medical School, Boston, MA
 \item Cancer Program, Broad Institute of Harvard and MIT, Cambridge, MA 
 \item Cancer Data Science Program, Dana-Farber Cancer Institute, Boston, MA
\item Contributed Equally
 \end{affiliations}
 
\noindent\textbf{Interactive Demo:} http://toad.mahmoodlab.org\\
 
\end{spacing}
\begin{spacing}{1.4}
\noindent\textbf{*Correspondence:}\\ 
Faisal Mahmood \\
60 Fenwood Road, Hale Building for Transformative Medicine\\
Brigham and Women's Hospital, Harvard Medical School\\
Boston, MA 02445\\
faisalmahmood@bwh.harvard.edu
\end{spacing}

\newpage
\noindent\textbf{\large{Abstract}}
\vspace{-10mm}
\begin{spacing}{1.38}
\noindent 




\noindent Cancer of unknown primary (CUP) is an enigmatic group of diagnoses where the primary anatomical site of tumor origin cannot be determined\cite{rassy2020progress,varadhachary2014cancer}. This poses a significant challenge, since modern therapeutics such as chemotherapy regimen and immune checkpoint inhibitors are specific to the primary tumor\cite{massard2011carcinomas}. Patients with a CUP diagnosis routinely undergo an extensive diagnostic work-up of pathology, radiology, endoscopy, laboratory tests, and clinical correlation in an attempt to determine the primary origin. Such exploration is not only time and resource consuming, but it might significantly delay administration of the suitable treatment. Despite extensive diagnostic work-ups the primary may never be determined in many cases. Recent work has focused on using genomics and transcriptomics for identification of tumor origins\cite{jiao2020deep, penson2020development, grewal2019application}. However, genomic testing is not conducted for every patient and lacks clinical penetration in low resource settings. Herein, to overcome these challenges, we present a deep learning-based computational pathology algorithm-TOAD-that can provide a differential diagnosis for CUP using routinely acquired histology slides. We used 17,486 gigapixel whole slide images with known primaries spread over 18 common origins to train a multi-task deep model to simultaneously identify the tumor as primary or metastatic and predict its site of origin. We tested our model on an internal test set of 4,932 cases with known primaries and achieved a top-1 accuracy of 0.84, a top-3 accuracy of 0.94 while on our external test set of 662 cases from 202 different hospitals, it achieved a top-1 and top-3 accuracy of 0.79 and 0.93 respectively. We further curated a dataset of 717 CUP cases from 151 different medical centers and identified a subset of 290 cases for which a differential diagnosis was assigned. Our model predictions resulted in concordance for 50\% of cases ($\kappa$=0.4 when adjusted for agreement by chance) and a top-3 agreement of 75\%. Our proposed method can be used as an assistive tool to assign differential diagnosis to complicated metastatic and CUP cases and could be used in conjunction with or in lieu of immunohistochemical analysis and extensive diagnostic work-ups to reduce the occurrence of CUP. 
\end{spacing}


\newpage
\begin{spacing}{1.35}
\vspace{-3mm}
For the vast majority of cancer diagnoses, the site of a primary tumor can be determined via pathological examination of tissue, or through a clinical and radiological assessment of the patient. However, 1-3\% of cases are often categorized as enigmatic cancers of unknown primary (CUP) where the anatomic site of primary origin cannot be assigned despite extensive diagnostic investigation and clinical correlation\cite{rassy2020progress,varadhachary2014cancer}. Decades of study have led to cancer treatment strategies that generally rely upon knowledge of the primary site of the tumor, whether it be surgical resection, radiation therapy, chemotherapeutic regimen, or targeted immunotherapies\cite{massard2011carcinomas}. 

\vspace{-5mm}
A majority of CUP cases where a putative primary cannot be assigned are treated with empirical chemotherapy and have poor prognosis (median survival 7-11 months, one year survival 25\%)\cite{rassy2020progress,varadhachary2014cancer}. Hence, CUP patients often undergo comprehensive diagnostic work-ups including pathology, radiology, endoscopic, and laboratory examinations to determine the occult primary site\cite{massard2011carcinomas,varadhachary2014cancer}. Recent work has proposed using molecular (genomic and transcriptomic) features for determining primary origin\cite{jiao2020deep, penson2020development,grewal2019application}. However, such testing is not routinely performed for every patient and lacks clinical penetration in low resource settings. The frontline of primary site classification remains tissue examination by a pathologist using histology with the aid of immunohistochemistry (IHC). However, despite the improvements from sophisticated imaging modalities, specific and sensitive immunohistochemical testing, and molecular profiling the diagnosis of CUP remains a current-day diagnostic challenge. Moreover, uncertainty in classifying a lesion as primary or metastatic and mistaking a relapse of an antecedent malignancy have also been reported in literature\cite{nass2009mir, estrella2011mucosal}. Evidence suggests that around 10\% CUPs can be prematurely diagnosed due to suboptimal investigation at the time of presentation\cite{rassy2020progress,varadhachary2014cancer}. Recent advances in deep learning\cite{esteva2019guide,lecun2015deep} have increasingly demonstrated accurate and reliable performance on a variety of different human identifiable features and phenotypes\cite{liu2020deep,lu2020data, campanella2019clinical,chen2019augmented,mei2020artificial,ouyang2020video,hollon2020near, esteva2017dermatologist, bulten2020automated,bera2019artificial} as well as phenotypes that are typically not recognized by human experts\cite{raghunath2020prediction,natmedlung,kather2019deep,tomavsev2019clinically,abduljabbar2020geospatial, poplin2018prediction, mitani2020detection}. 


\vspace{-6mm}
\section*{Tumor origin assessment via deep learning}
\vspace{-6mm}
In order to address the difficulties and complexities associated with identifying the primary sites of tumor specimens, we propose a deep learning-based solution that uses scanned H\&E whole slide images (WSIs), which are routinely used for clinical diagnosis, for identifying the site of primary origin without immunohistochemical analysis, genomic testing, or extensive clinical diagnostic screening. We developed Tumor Origin Assessment via Deep-learning (TOAD), a high-throughput, interpretable deep learning framework that can be used to simultaneously predict whether the histological sample is metastatic and assign differential diagnosis for primary origin. In addition to addressing an unmet need in the diagnosis of CUP patients, TOAD can also act as an assistive tool for pathologists for complicated metastatic cases where a large number of IHCs are required to narrow a differential diagnosis. TOAD is capable of providing assistance with differential diagnosis (top-3, top-5 predictions) instead of a single diagnosis for the pathologists consideration. Such differential diagnosis are a routine part of the clinical and pathological work-up for CUP cases and assist with narrowing down possibilities of potential primaries. Our study uses 24,885 WSIs from 23,273 patient cases from the Brigham and Women's Hospital and the TCGA, where each slide was treated as an independent case. We trained our model using 17,486 WSIs using our weakly-supervised multi-task training paradigm. Then, extensive analysis was conducted to assess the performance of TOAD by first testing on 4,932 WSIs with known primaries, and carefully analyzing complicated metastatic cases to determine the capability of TOAD for assigning differential diagnosis. Second, to further assess the adaptability of our model we evaluated on an external multi-institutional test set of 662 cases from 202 different medical centers. Third, we curated an additional test dataset of 717 consented CUP cases received from 151 medical centers that could not be assigned a primary using histology alone and identified a subset of 290 cases where a primary differential was identified based on immunohistochemical analysis, radiology, patient history, clinical correlation or at autopsy (see \textbf{Extended Data Figure 1} for an overview of our study design). 

\begin{figure*}
\vspace{-12mm}
\includegraphics[width=\textwidth]{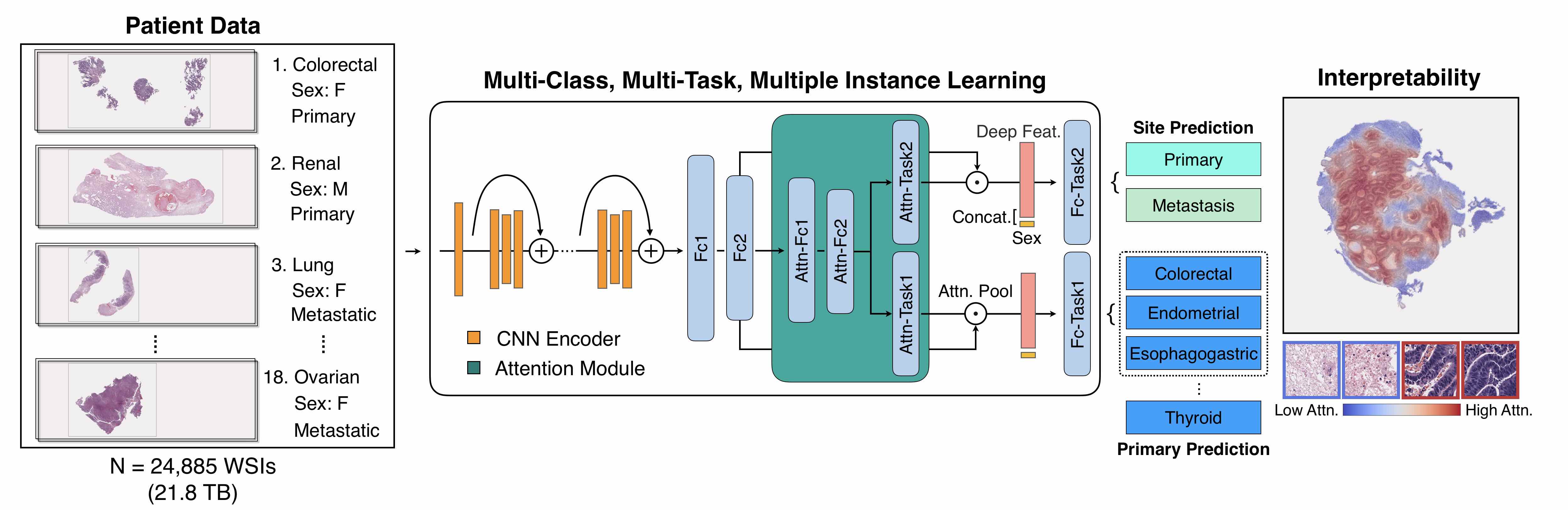}
\caption{\textbf{Tumor Origin Assessment via Deep Learning (TOAD) workflow.} Patient data in the form of digitized high-resolution FFPE H\&E histology slides (known as WSIs) serve as input into the main network. For each WSI, the tissue content is automatically segmented and divided into an average of thousands to tens of thousands of regions as small image patches. These images are processed by a pretrained convolutional neural network, which serves as an encoder to extract a compact, descriptive feature vector from each patch. Using an attention-based multiple instance learning algorithm, TOAD learns to rank all tissue regions in the slide using their feature vectors and aggregate their information across the whole slide based on their relative importance, assigning greater weights to regions perceived to have high diagnostic relevance. As an additional covariate, the patient's sex can be fused with the aggregated histology features to further guide classification. By using a multi-branched network architecture and a multi-task objective, TOAD can predict both the tumor origin as well as whether the cancer is primary or metastatic. Additionally, the attention scores that the network assigns to each region can be used to interpret the model's prediction.} 
\vspace{-2mm}
\end{figure*}

\vspace{-4mm}
Our weakly-supervised multi-task deep learning classifier model was trained on gigapixel WSIs without requiring manual expert annotation for regions of interest (ROIs) and predicts major primary sites at the slide level. We combine transfer learning and weakly-supervised multi-task learning to allow a single, unified predictive model to be efficiently trained on tens of thousands of WSIs, a scale that is likely required to solve both the complex problem of classifying 18 common cancer origins and predicting if the cancer is metastatic or primary simultaneously.
Using attention-based learning, our approach automatically learns to locate regions in the slide that are of high diagnostic relevance and aggregates their information to make the final predictions. Subsequently, using custom visualization tools, the relative importance of each region examined by the model can be intuitively displayed as high-resolution attention heatmaps for human interpretability and validation (\textbf{Figure 3, Extended Data Figure 6, Interactive Demo}).

\vspace{-4mm}
TOAD begins by automatically segmenting and patching the tissue regions in the WSI into many smaller cropped regions that can be directly processed by a convolutional neural network (CNN). Using transfer learning, a deep residual CNN is first deployed as an encoder to compress the raw input data by embedding them into compact low-dimensional feature vectors for efficient training and inference. Following feature extraction, TOAD uses a custom, lightweight neural network that takes in the deep features of all tissue regions in the slide as input for weakly-supervised learning. Building upon the attention-based pooling operator\cite{ilse2018attention,lu2020data}, an attention module learns to rank each region's relative importance toward the determination of each classification task of interest and aggregates their deep feature representations into a single slide-level feature vector for each task by computing their attention-score-weighted average. Further, we explored  incorporating the patient's gender as an additional covariate by fusing it with the slide-level features via concatenation before the final classification layers that predict independently both the cancer origin (multi-class classification) and whether the tumor is primary or metastatic (binary classification). The two classification problems are learned jointly during training by using a multi-task objective and sharing the model parameters of intermediate layers. A separate attention branch layer, however, is used for each task to increase the model's expressivity, allowing it to attend to different sets of information-rich regions of the slide depending on the task (\textbf{Figure 1}). Further details of the model architecture, training and dataset are described in the methods and \textbf{Extended Data Figure 1, Extended Data Table 1}.

\end{spacing}

\begin{figure*}
\vspace{-12mm}
\includegraphics[width=\textwidth]{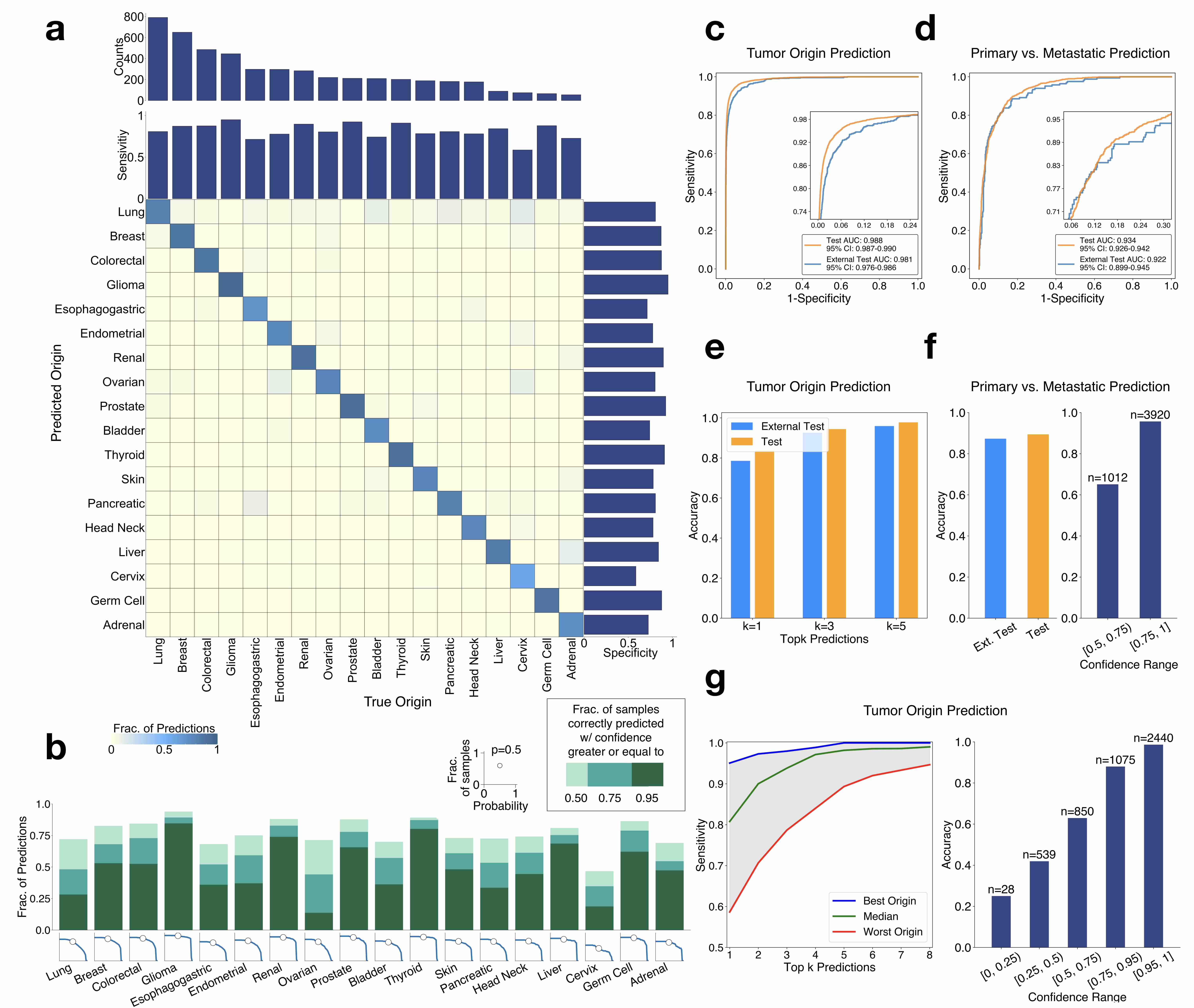}
\caption{\textbf{Model performance of TOAD.} \textbf{a.} Slide-level classification performance on the test set (n=4,932) for 18 tumor origins. Columns represent the tumor's true origin and rows represent the model's predicted origins. \textbf{b.} For each origin, fraction of samples correctly classified with a confidence (probability) score of greater than 0.5, 0.75 and 0.95 respectively (\textbf{top}); fraction of samples (y-axis) correctly classified at or above a certain confidence threshold (x-axis, computed over increments of 0.025 in probability score) (\textbf{bottom}). \textbf{c.} Micro-averaged received operator characteristic (ROC) curves for the multi-class classification of the tumor origin, evaluated on the test set (n=4,932) and an independent test set of external cases only (n=662). \textbf{d.} ROC curves for the auxiliary task of predicting primary vs. metastasis in the test set and external test set. \textbf{e.} Top-k accuracy of model for tumor origin classification on the test set and external test set for k $\in \{1,3,5\}$. \textbf{f.} Overall accuracy (\textbf{left}) of model for predicting primary vs. metastasis and \textbf{right.} accuracy of model's predictions stratified into low confidence ($0.5 \leq p < 0.75$) and high confidence ($0.75 \leq p \leq 1.0$). \textbf{g. left.} Sensitivity score for the best, median and worst tumor origin for based on the model's top-k predictions for k $\in \{1,2,...,8\}$. \textbf{right.} Accuracy of predictions for different bins of prediction confidence.} 
\end{figure*}
\begin{spacing}{1.45}
\vspace{-8mm}
\section*{Evaluation of model performance}
\vspace{-6mm}
We evaluated our proposed deep learning framework by partitioning our dataset, which consists of a total of 24,885 FFPE H\&E digitized diagnostic slides from 23,297 patient cases, into 70/10/20 splits for training, validation and testing, respectively. On this held-out test set of 4,932 slides with known primaries that were not previously seen by the model, TOAD achieved an overall accuracy of 83.6\%, and a micro-averaged AUC ROC of 0.988 (95\% CI: 0.987 - 0.990) (\textbf{Figure 2c}). When the model is evaluated using top-k differential diagnosis accuracy, \textit{i.e}, how often the groundtruth label is found in the model's k highest confidence predictions, TOAD achieved a top-3 accuracy of 94.4\% and top-5 accuracy of 97.8\% (\textbf{Figure 2e}). Such top differential predictions can be extensively useful for complicated metastatic and CUP cases where narrowing down potential primaries can assist with the diagnostic workflow and reduce the number of IHC stains and other diagnostic tests required to pin a culprit primary. \textbf{Figure 2a} shows performance for each individual primary site and a full summary table of classification performance metrics including precision, recall, F1-score, and one-vs-rest AUC ROC are included in \textbf{Extended Data Table 3}. The training with validation performance over time for models of different configurations are shown in \textbf{Extended Data Figure 5} (see \textbf{Supplementary Data File Table 1} for individual case assessments). By binning the model's predictions based on their confidence, we note that the majority of predictions on the test set are made with high confidence, \textit{e.g.} 2,440/4,932 predictions have a confidence of 0.95 or higher, and this bin of predictions also resulted in the highest accuracy of 98.5\% compared to less confident predictions (\textbf{Figure 2b}). Together, the high top-k accuracy suggests that we can potentially use TOAD's top predictions for a given slide to narrow down the origin of the tumor to a handful of possible locations while predictions with high confidence (\textit{e.g.} $\geq 0.75$) are generally reliable (\textbf{Figure 2g}). For interpretability and further validation, we examined attention heatmaps for tumors metastasized from the lung, breast and colon and confirmed that high attention regions generally exhibit tumor morphology characteristic of the respective primary tumor (\textbf{Figure 3}, \textbf{Extended Data Figure 6}). Additionally, TOAD was able to predict whether the tumor specimen is a primary or metastatic tumor with an accuracy of 89.4\% and an AUC ROC of 0.934 (95\% CI: 0.926 - 0.942) (\textbf{Figure 2d}), high confidence predictions were assigned to the majority of correct cases (\textbf{Figure 2f}). \textbf{Extended Data Table 7} shows site-wise performance for this binary task. 
\end{spacing}

\begin{spacing}{1.40}
\vspace{-8mm}
\section*{Generalization to multi-institutional external test cohort}
\vspace{-6mm}
To assess the adaptability of our model across different healthcare systems with different H\&E staining protocols and patient populations, we also validated TOAD on an additional test set of 662 external cases submitted from 202 US and international medical centers (for geographic diversity, see \textbf{Extended Data Figure 4, Supplementary Data File Table 4}). Without tuning or any form of domain adaptation, our trained model produced an accuracy of 78.5\%, top-3 accuracy of 92.6\%, top-5 accuracy of 95.9\% and AUC ROC of 0.981 (95\% CI: 0.976 - 0.986) on this additional independent test set (\textbf{Figure 2c,e}). Similarly, on the second task of distinguishing between metastasis and primary tumor, the model scored an AUC of 0.922 (95 \% CI: 0.899 - 0.945) and accuracy of 87.3\% (\textbf{Figure 2d,f}). The model's performance is consistent with results on the first test set, indicating that our model is capable of generalization to diverse data sources not encountered during training. Individual cases assessments are available in \textbf{Supplementary Data File Table 2.}


\end{spacing}

\begin{figure*}
\vspace{-12mm}
\includegraphics[width=\textwidth]{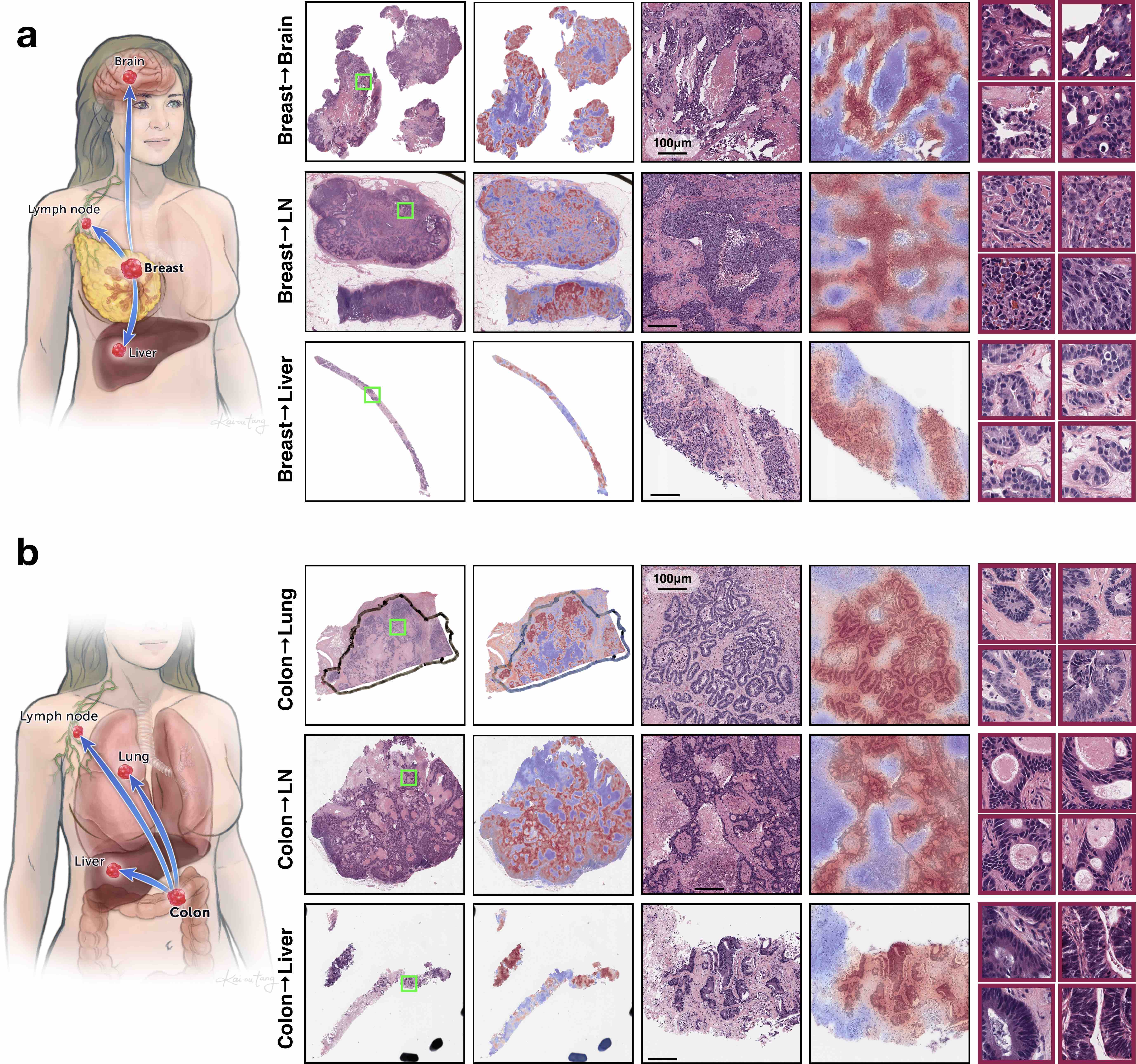}
\caption{\textbf{Exemplars of metastases from primary sites with attention heatmaps}. For all cases, smooth attention scores are computed on overlapping 256 x  256 patches, normalized using percentiles and displayed on top of the original H\&E WSI as a semi-transparent overlay where overlaid regions range from crimson (high attention, high diagnostic relevance) to navy (low attention, low diagnostic relevance). From left to right, low magnification with corresponding attention map, medium magnification with corresponding attention map, and high magnification patches of high attention regions. \textbf{a}: Medium and high magnification views demonstrate sheets of cells as well as small tubules and glands, morphologies consistent with metastatic breast carcinomas. \textbf{b}: Medium and high magnification views demonstrate so-called "dirty necrosis" and variably-sized glands with densely-packed, hyperchromatic nuclei, characteristic of colorectal adenocarcinoma. The attention heatmaps allow the model's predictions for each case to be visually interpretible for human experts, revealing the morphological features used by the model for classification determination. By studying attention heatmaps for different metastatic tumors, we verified that the model is attending strongly to tumor regions for predicting the site of origin as expected. More attention heatmaps for tumors metastasized from the lung are shown in \textbf{Extended Data Figure 6} and high resolution heatmaps for cases from all primary sites can be accessed through our interactive demo available at http://toad.mahmoodlab.org.}  
\end{figure*}

\begin{table}
\vspace{-16mm}
\caption*{\textbf{Table 1. Testing on primary, metastasis of known primary and CUP}}
\centering
\subcaption*{\textbf{A. Performance on Overall Test and External Test Set}}
\begin{tabular}{p{8cm}p{2cm}p{2cm}p{2cm}p{2cm}}
\toprule
Primary and Met. Test Set   & Cohen's $\kappa$ & Top-1 Acc & Top-3 Acc & Top-5 Acc \\
\midrule
Test Cases (n=4932)         & 0.820         & 0.836     & 0.944     & 0.978     \\
External Test Cases (n=662) & 0.746         & 0.785     & 0.926     & 0.959    \\
\bottomrule
\end{tabular}
\bigskip

\subcaption*{\textbf{B. Performance Analysis on Challenging Metastatic Cases}}
\begin{tabular}[Hht!]{p{8cm}p{2cm}p{2cm}p{2cm}p{2cm}}
\toprule
Metastatic Test Set              &   Cohen's $\kappa$ & Top-1 Acc & Top-3 Acc & Top-5 Acc \\
\midrule
All Cases (n=882)                     & 0.567 & 0.626      & 0.849     & 0.923     \\
\midrule
Required no IHC (n=264)               & 0.610 & 0.659     & 0.898     & 0.939     \\
Required between 1 and 5 IHC (n=303) & 0.551 & 0.617     & 0.838     & 0.911     \\
Required $\ge$5 IHC (n=200) & 0.470 & 0.560     & 0.790     & 0.915     \\
\midrule
Clinical correlation recommended (n=64)       & 0.515 & 0.609     & 0.766     & 0.938     \\
Poorly differentiated (n=155)          & 0.475 & 0.587     & 0.839     & 0.929 \\
\bottomrule
\end{tabular}
\bigskip

\subcaption*{\textbf{C. Performance on CUP Cases}}
\begin{tabular}{p{5cm}p{2cm}p{3cm}p{3cm}p{3cm}}
\toprule
CUP Test Set & Cohen's $\kappa$ & Top-1 Agreement & Top-3 Agreement & Top-5 Agreement \\
\midrule
Primary assigned (n=290)               & 0.397         & 0.500           & 0.745           & 0.900           \\
Confidence $\ge$ 0.5 (n=196) & 0.512         & 0.597           & 0.806           & 0.913           \\
Confidence $\ge$ 0.9 (n=53)  & 0.705         & 0.755           & 0.868           & 0.906           \\ \midrule
High certainty (n=185)                 & 0.476         & 0.573           & 0.816           & 0.957           \\
Low certainty (n=105)                  & 0.261         & 0.371           & 0.619           & 0.800          \\ \bottomrule
\end{tabular}
\vspace{-3mm}
\end{table}

\begin{spacing}{1.42}
\vspace{-7mm}
\section*{Evaluation on challenging metastatic cases}
\vspace{-7mm}
It is challenging to objectively evaluate the model's ability to correctly predict the origin of tumor for CUP cases because there are limited and weak ground truth labels. In light of this challenge, we first analyzed the performance of TOAD on difficult metastatic cases in our test set for which a diagnosis is available. On these 882 metastatic cases, TOAD achieved a micro-averaged AUC of 0.939 (95\% CI: 0.930 - 0.948) and overall accuracy of 62.6\%, top-3 accuracy of 84.9\% and top-5 accuracy of 92.3\% (\textbf{Table 1. B}). This demonstrates that TOAD can assist with assigning a differential diagnosis by narrowing down possible origins. The sensitivity for correctly identifying these cases as metastatic, using the prediction from our multi-task network is 70.9\%. Furthermore, we queried the pathology report for each metastatic case in our database and were able to extract the number of IHC tests performed in 767 reports (median: 2, min: 0, max: 27). Using the number of IHC tests performed as an indirect measure of the difficulty in diagnosing the case, we examined the performance of TOAD across different levels of IHC usage. 
As expected, in cases that were diagnosed without requiring IHC ($n=264$), TOAD scored the highest accuracy of 65.9\%, and a top-3 and top-5 accuracy of 89.8\% and 93.9\% respectively. However, even in the more difficult cases ($n=200$) that required 5 (75th percentile) or more IHC tests, TOAD still managed to achieve an accuracy of 56.0\%, and a top-3 accuracy of 79.0\%. Moreover, the top-5 accuracy for these difficult cases remained at 91.5\%. Similarly, we identified and analyzed performance on two other subsets of challenging cases, including 64 cases that could not be diagnosed with IHC analysis and required further clinical or radiologic correlation to make the diagnosis (top-1: 60.9\%, top-3: 76.6\%, top-5: 93.8\%) and 155 cases that were characterized as poorly-differentiated tumor in the pathology reports (top-1: 58.7\%, top-3: 83.9\%, top-5: 92.9\%). As expected, performance for challenging and difficult to diagnose cases were lower, but largely consistent with performance on the entire set of metastatic cases. It is worth noting that the model was able to achieve this level of performance without having access to additional clinical variables, or IHC results, as it makes its predictions solely based on the digitized H\&E slide and the patient's gender. Overall, these results again suggest the potential of TOAD to provide reliable and valuable candidate primaries based its top predictions to guide further differential diagnosis, IHC work-up and reduce the requirement of laboratory tests and clinical correlation. We also calculated Cohen's kappa\cite{mchugh2012interrater}, which measures the inter-observer agreement between the model and the assigned differential, while taking into account agreement by chance. The $\kappa$ scores fell in the range of moderate to substantial agreement for metastatic cases and indicated even stronger agreement on our overall test and external test sets (\textbf{Table 1. A, B.}). 

\end{spacing}

\begin{spacing}{1.32}
\vspace{-6mm}
\section*{Evaluation on a multi-institutional CUP cohort}
\vspace{-6mm}
\noindent We further curated a dataset of 717 consented cases from 151 medical centers that were assigned a diagnosis of CUP at some point during their course of diagnosis and treatment. These challenging cases were submitted from 146 US and 5 international medical centers (\textbf{Extended Data Figure 4, Supplementary Data File Table 5)}. None of these cases could be assigned a primary diagnosis using the histology slide alone. Instead, all cases required thorough IHC testing and the patients underwent extensive clinical workups (including radiology, endoscopy, etc.) in an attempt to determine the occult primary anatomic site. For a more thorough evaluation of our model, we carefully analyzed all available electronic medical records (EMRs) for patients in this CUP dataset, including clinical and familial history, radiology reports, endoscopy reports, treatment, and follow up history. We identified a subset of 290 cases which were assigned a primary differential at some point during the course of diagnosis and treatment, while the remaining cases could not be assigned a primary at any point or had limited medical records. As expected, these differential diagnosis for CUP cases involve elements of uncertainty and conjecture and should be distinguished from confident, ground truth labels, which cannot be realistically obtained for CUP cases. 

\vspace{-5mm}

We used our trained TOAD model to assign predictions for each case in our CUP dataset (using only information contained in the histology slide and the patient's gender) and observed that the model's top prediction directly concorded with the site indicated by the primary differential assigned in 145 of the 290 cases (50.0\%), with a $\kappa$ score of 0.397, indicating that there was fair agreement by chance between the assigned differentials and the model's predictions (\textbf{Table 1. C}, see \textbf{Supplementary Data File Table 3} for individual case predictions). When using the model's top-3 and top-5 predictions, the agreement jumps to 74.5\% and 90.0\% of cases respectively. This is a particularly encouraging result since our model was able to assign concordant differential diagnosis based on the histology image, such differentials are typically assigned using extensive investigative diagnostic work-ups. We observed that similar to what we found on the test sets, a significant fraction of the model's predictions were made with high confidence (\textit{e.g.} 196 out of 290 predictions had a confidence score of $\ge$0.5). We noted higher agreement in these high confidence predictions; encouragingly, among the 53 predictions made with a confidence score of 0.9 or higher, we noted substantial agreement between the model and the primary differential ($\kappa=0.705$). This serves as further evidence for our hypothesis that the model's high confidence predictions are generally more reliable, which we also observed on our test set of confirmed metastatic and primary cases. Here, the model's average confidence on all 717 CUP cases is 0.605 (median: 0.579) and 457/717 (63.8\%) of cases are predicted with a confidence of 0.5 or higher.

\vspace{-5mm}

We further subcategorized the CUP cases into high-certainty diagnoses (n=185) and low-certainty (n=105) based on the strength of the evidence used to make the determination, language used in EMRs and whether the cancer was treated based on a certain primary as well as if the patient responded to that particular treatment. As expected, agreement is poor for cases in the low-certainty bin ($\kappa=0.261$) while for high-certainty diagnoses, higher agreement was observed ($\kappa=0.476$) across all metrics (\textbf{Table 1C}). In \textbf{Extended Data Figure 8 and 9}, we demonstrate in two independent cases, how the top predictions from the TOAD model can be used in conjunction with IHC testing to assist in suggesting and determining the origins of challenging metastatic cases initially assigned a diagnosis of CUP.


\vspace{-4mm}

\end{spacing}

\begin{spacing}{1.42}
\vspace{-6mm}
\section*{\large{Discussion}}
\vspace{-4mm}
In this study, we presented TOAD, the first deep learning algorithm developed to predict the origin of a tumor based on whole-slide histopathology. TOAD targets the difficult problem of assigning origins for cancers of unknown primary using just histology WSIs, a task that is typically accomplished using extensive clinical work-ups including IHC analysis, radiologic imaging and clinical correlation. By using weakly-supervised learning, our algorithm was developed using tens of thousands of cases without requiring any fine-grained manual annotation of the slide or representative ROIs and can be easily tested on digitized histology images of arbitrary size. To demonstrate the effectiveness of our proposed algorithm, we trained and validated our model using a large dataset of diagnostic WSIs and showed consistent performance between a large test set of metastatic and primary cases and a geographically diverse, independent test set of external cases received from over two hundred different institutions. 
It has been shown that pathologists show limited ability to identify the origins of metastatic tumors when provided with minimal clinical information and especially when evaluating based on morphology alone\cite{sheahan1993metastatic}. We show that despite using only histology and the patient's gender as input for decision making, our model can make fairly accurate predictions particularly in assigning top-3 or top-5 primary differentials even for challenging metastatic cases that required extensive IHC tests and clinical or radiologic correlation to diagnose. 
Lastly, we also curated a large set of CUP cases, which were acquired from a diverse cohort of US and international medical centers, and we subsequently identified a subset of cases for which a primary differential was assigned at some point after the initial diagnosis, often after extensive clinical workups. For these extremely challenging cases, the H\&E slide proved to be insufficient for human experts to assign a primary whereas despite being limited to just the patient's gender and morphological information in the WSIs, our model was able to make predictions that are concordant with the primary differentials assigned after IHC work-up to a meaningful degree. 
We also showed that metadata such as the patient's gender can be incorporated in the model and using multitask learning, we can use a single model to additionally predict whether a tumor is primary or metastatic without sacrificing performance. We conducted ablation experiments to investigate the effect of adding gender as a covariate and multi-task learning and found minimal difference in performance (\textbf{Extended Data Table 4}).

\vspace{-4mm}
As further analysis, we also showed that our multi-task network can distinguish between primary tumors and metastatic tumors found at the same site, which can pose occasional difficulties to pathologists\cite{nass2009mir, estrella2011mucosal}. As an example, when asked to predict tumors found in the central nervous system (CNS) as primary or metastatic, the model reached an accuracy of 95.0\% ($n=446$) on gliomas and other tumors metastasized to the brain and similarly reported high accuracy for GI metastatic sites (colorectal: 92.9\% accuracy for $n=406$ and esophagogastric: 94.1\% accuracy for $n=273$) in our test set (\textbf{Extended Data Figure 7}). From additional experiments, we confirmed that TOAD can also be applied to predict the primary for subsets of tumors that share the same morphological appearance (\textit{e.g.} various subtypes of squamous cell carcinoma, \textbf{Extended Data Figure 2}) or have metastasized to a common site (\textit{e.g.} the lymph node, \textbf{Extended Data Figure 3}). While more restricted in their scope than our main 18-class classifier, these  networks targeting specific subgroups of tumors have the potential to serve as additional readers when attempting to rule out plausible candidate origins proposed by the main network (\textbf{Extended Data Figure 9}).

\vspace{-4mm}
Overall, an encouraging observation is that a substantial fraction of our model's predictions were made with high confidence, which also consistently proved to be more reliable and accurate. By using the top-k predictions, the model is also capable of narrowing the tumor origin down to a handful of possible locations (\textit{e.g.} top-3 or 5 most likely locations) with fairly high accuracy. This suggests the potential clinical applicability of TOAD to both suggest high-likelihood candidate primaries when the diagnosis of a case is initially ambiguous and as a second reader to human experts, potentially prompting re-evaluation or exploring alternative hypotheses when the model produces a high confidence prediction in disagreement. In such cases, the attention heatmap and high attention patches (\textbf{Figure 3, Extended Data Figure 6, Interactive Demo}) may be used in conjunction with the model's probability score predictions for human interpretability and validation. Unlike previous works that predicted the cancer type based on genomic alterations in the tumor, as the first histology-based deep learning algorithm proposed and validated for automated prediction of tumor origins, our approach is also arguably more broadly applicable, especially for low-resource settings where clinical expertise, immunohistochemistry and molecular testing may be limited.
\end{spacing}

\vspace{-5mm}
\section*{\large{Online Methods}}
\begin{spacing}{1.5}
\vspace{-4mm}
\noindent\textbf{Dataset Description} \\
For model development, we curated a dataset of 14,518 WSIs from internal consented patient cases at the Brigham and Women's Hospital. These slides were collected between 2010 - 2019 and scanned at 20$\times$ using an Aperio scanner. Unless indicated otherwise, each WSI corresponds to a unique patient. We grouped these cases into 18 common cancer origins, where each origin encompasses both common and rare tumor subtypes for which at least 10 cases were found in our database (\textbf{Extended Data Table 2}). All data used for this study were anonymised. We additionally queried the TCGA Data Commons and all diagnostic WSIs from repositories corresponding the 18 classes. Among slides downloaded from the TCGA, slides which do not contain tumors or that lacked lower magnification downsamples were excluded. In total, we gathered 10,367 WSIs from 8,755 patient cases across the 24 TCGA studies. Our overall dataset was composed of 24,885 FFPE H\&E digitized diagnostic slides (20,413 primary and 4,472 metastatic WSIs from 23,297 patient cases; 54.7\% F, 45.3\% M) (\textbf{Extended Data Table 1}). This roughly amounted to 21.8 Terabytes of raw data. This dataset is randomly partitioned and is stratified by class, into a training set (70\% of cases), a validation set (10\% of cases) and a test set (20\% of cases). The partitioning was performed at the patient-level and therefore all slides from the same patient are always placed into the same set. Additionally, we processed 662 external consult cases consented for research and received at the Brigham \& Women's Hospital from 202 medical centers across 34 states in USA and 19 international medical centers from 8 other countries (see \textbf{Extended Data Figure 4} for geographic diversity). Slides for these cases were prepared at their respective institutions using a variety of different tissue preparation, processing and staining protocols. A full origin-wise breakdown of these datasets are summarized in \textbf{Extended Data Table 1}. Lastly, to further validate our model, we also identified 717 consented cases that were assigned a diagnosis of CUP. These cases are received from 146 medical centers across 22 US states and 5 international centers from 2 other countries (see \textbf{Extended Data Figure 4} for geographic diversity). For each case we reviewed electronic medical records including pathology report in combination with laboratory results, patient history, oncology, radiology, endoscopy and autopsy reports where applicable and if available, we determined a subset of 290 cases with a primary differential. These differentials were assigned during the course of diagnosis or treatment. It was verified that none of these cases could be diagnosed using histology alone and required extensive immunohistochemical analysis or clinical correlation. While it is not possible to obtain a ground truth for CUP cases such analysis based on the differential diagnosis was used to assess the value of our model in assigning appropriate differentials to CUP cases using histology alone. For further analysis, we additional split these 290 cases into high-certanity (n=185) and low-certainty differentials (n=105) based on the language and evidence used in the electronic medical records (\textbf{Extended Data Figure 1} for detailed study design). 

\noindent\textbf{Multi-task Weakly-Supervised Computational Pathology} \\
We used deep learning to simultaneously predict the origin of tumor in each WSI and whether it is the primary site or a metastasis. Due to the enormous size of gigapixel WSIs as well as the large variation in the shape of tissue content captured by the image, it is generally considered inefficient, unintuitive, and intractable to deploy deep learning algorithms based on convolutional neural networks (CNN) directly on top of the entire WSI for training or inference. While it is possible to use smaller regions of interests (ROIs) for training, this approach has the drawback that since the slide-level diagnosis (\textit{e.g.} Lung Adenocarcinoma) is only manifested in a fraction of the tissue content in the WSI, unless human expertise and manual labor is involved to ensure these smaller regions are representative of the diagnosis made for the entire slide, naively associating them with the slide-level diagnosis will lead to noisy and erroneous labels. To overcome this limitation, we built a compact neural network model and used a form of weakly-supervised machine learning known as multiple instance learning. By considering each WSI as a collection (known as a bag) of smaller image regions (known as instances), we trained the multi-task network directly with slide-level labels without the need for manually extraction of regions of interests (ROIs), while also taking into account information from the entire slide. \\
For computational efficiency, we first performed dimensionality reduction on the raw image data by encoding each 256 $\times$ 256 RGB image patch into a 1024-dimensional feature vector using a pretrained CNN (transfer learning). In the low-dimensional feature space, the information from all tissue regions in each slide are aggregated by extending attention-based pooling\cite{ilse2018attention} to multiple tasks, based on which the classification layers of the network outputs the final slide-level predictions. 
Specifically, two stacked fully-connected layers Fc\textsubscript{1} and Fc\textsubscript{2}, parameterized by $\mathbf{W}_{1} \in \mathbbm{R}^{512 \times 1024}, \mathbf{b}_{1} \in \mathbbm{R}^{512}$ and $\mathbf{W}_{2} \in \mathbbm{R}^{512 \times 512}, \mathbf{b}_{2} \in \mathbbm{R}^{512}$ in the base of the network allow the model to learn histology-specific feature representations by tuning deep features extracted through transfer learning, mapping the set of patch feature embeddings $\{\mathbf{z}_{k}\} \in \mathbbm{R}^{1024}$ in a given WSI to 512-dimensional vectors: 
\begin{equation} \label{fc}
    \mathbf{h}_{k} = \text{ReLU}(\mathbf{W}_{2}(\text{ReLU}(\mathbf{W}_{1}\mathbf{z}_{k}^{\top} + \mathbf{b}_{1})) + \mathbf{b}_{2})
\end{equation}
\noindent \textit{Multi-task Attention Pooling.}
In the proposed multi-task learning framework, the multi-layered attention module consists of layers Attn-Fc\textsubscript{1} and Attn-Fc\textsubscript{2} with weight parameters $\mathbf{V}_{a} \in \mathbbm{R}^{384 \times 512}, \text{and } \mathbf{U}_{a} \in \mathbbm{R}^{384 \times 512}$ (shared across all tasks), and one independent layer $\mathbf{W}_{a, t} \in \mathbbm{R}^{512 \times 384}$ for each task $t$. This network module is trained to assign an attention score $a_{k,t}$ (eqn \ref{attention}) to each patch, where after Softmax activation, a high score (near 1) indicates that a region is highly informative towards determining the slide-level classification task and a low score (near 0) indicates the region has no diagnostic value (for simplicity the bias parameters are not shown in the equation):
\begin{equation} \label{attention}
a_{k, t}=\frac{\exp \left\{\mathbf{W}_{a, t}\left(\tanh \left(\mathbf{V}_{a} \mathbf{z}_{k}^{\top}\right) \odot \operatorname{sigm}\left(\mathbf{U}_{a} \mathbf{z}_{k}^{\top}\right)\right)\right\}}{\sum_{j=1}^{N} \exp \left\{\mathbf{W}_{a, t} \left(\tanh \left(\mathbf{V}_{a} \mathbf{z}_{j}^{\top}\right) \odot \operatorname{sigm}\left(\mathbf{U}_{a}\mathbf{z}_{j}^{\top}\right)\right)\right\}}
\end{equation}
Attention pooling then simply averages the feature representations $\{\mathbf{h}_{k}\}$ of all patches in the slide, weighted by their respective predicted attention scores $\{\mathbf{a}_{k, t}\}$, and the resulting feature vector $\mathbf{h}_{slide,t} \in \mathbbm{R}^{512}$ is treated as the histology deep features representing the entire slide for task $t$. This intuitive, trainable aggregation function allowed the network to learn to automatically identify the subset of informative regions in the slide in order to predict the primary without requiring detailed annotation outlining the precise regions of tumor. \\
\noindent \textit{Late-stage Fusion and Classification.}
We adopt a simple fusion mechanism to incorporate a patient's biological sex into the model's prediction by treating the sex $s$, as an additional covariate encoded by binary values, and concatenating it to the deep features extracted from the histology slide. The concatenation results in a 513-dimensional feature vector that is fed into the final classification layer $\mathbf{W}_{cls, t}$ for task $t$ to obtain the slide-level probability prediction scores:
\begin{equation}
    \mathbf{p}_t = \text{Softmax}(\mathbf{W}_{cls, t}\text{Concat}([\mathbf{h}_{slide,t}, s]) + \mathbf{b}_{cls, t})
\end{equation}
In our study, the first task of predicting the origin site of tumor is a 18-class classification problem and the second task of predicting whether a tumor is primary or metastatic is a binary problem. Accordingly, the task-specific classfication layers are parameterized by $\mathbf{W}_{cls, 1} \in \mathbbm{R}^{18 \times 513}$ and $\mathbf{W}_{cls, 2} \in \mathbbm{R}^{2 \times 513}$ respectively. \\
\noindent \textit{Training Details.} We randomly sampled slides using a mini-batch size of 1 WSI and used multi-task learning to supervise the neural network during training. For each slide, the total loss is a weighted sum of loss incurred from the first task of predicting the tumor origin and the loss from the second task of predicting primary vs. metastasis:
\begin{equation}
\mathcal{L}_{total} = c_1\mathcal{L}_{cls,1} + c_2\mathcal{L}_{cls,2}
\end{equation}
The standard cross-entropy was used for both tasks and to give higher importance to the main task of tumor origin prediction, we used $c_1=0.75$ and $c_2=0.25$.
After each mini-batch, the model parameters are updated via the Adam optimizer with an L2 weight decay of 1e-5 and a learning rate of 2e-4. To curb the model from potential over-fitting, we also used dropout layers with $p=0.25$ after every hidden layer. \\
\noindent\textit{Model Selection.}
During training, the model's performance on the validation set was monitored each epoch. Beyond epoch 50, if the validation loss on the tumor origin prediction task had not decreased for 20 consecutive epochs, early stopping was triggered and the best model with the lowest validation loss was used for reporting the performance on the held-out test set. \\
\noindent\textbf{Additional Experiments} \\
\textit{Classification of Adenocarcinoma and Squamous Cell Carcinoma.} \\
The Adenocarcinoma model was developed using a subset of 8292 adenocarcinoma WSIs that fall under 5 of the 18 tumor origin classes considered by the main network: Lung (2558), Colorectal (2448), Esophagogastric (1320), Prostate (1101) and Pancreatic (865). Similarly, the Squamous Cell Carcinoma (SCC) network was developed using a subset of 1707 SCC WSIs from 4 origins: Lung (854), Head Neck (424), Cervix (264), and Esophagogastric (165). 
For all experiments, the cases were partitioned into 70/10/20 splits for training/validation/testing. The model architecture, learning schedule and hyperparameters used were the same as for the main network. \\
\textit{Classification of tumor metastasized to the liver and lymph.} \\
The lymph site-specific model was developed using a subset of 697 WSIs of metastatic tumors from four primary origins including: Lung (341), Breast (185), Skin (110) and Thyroid (61). The liver site-specific network was developed using a subset of 740 WSIs of metastatic tumors from four primary origins including: Pancreatic (225), Colorectal (224), Breast (179), and Lung (112). For all experiments, the cases were partitioned into 70/10/20 splits for training/validation/testing. The model architecture, learning schedule and hyperparameters used were the same as for the main network except the mulitask attention branch for predicting primary vs. metastatic was disabled since all cases were metastatic. \\
\noindent\textbf{Computational Hardware and Software} \\
We processed all WSIs on Intel Xeon multi-core CPUs (Central Processing Units) and a total of 16 NVIDIA P100 and 2080 Ti GPUs (Graphics Processing Units) using our custom, publicly available CLAM\cite{lu2020data} whole slide processing pipeline implemented in Python. Each deep learning model was trained on multiple GPUs using the Pytorch deep learning library (version 1.5). Unless otherwise specified, plots were generated in Python (version 3.7.5) using matplotlib (version 3.1.1) and numpy (version 1.18.1) was used for vectorized numerical computation. The geographic diversity maps were generated using additional Python packages including pyshp (version 2.1.0), basemap (version 1.1.0) and geopy version (version 1.22.0). The confusion matrix plot was created in R (version 3.6.3) using ComplexHeatmap (version 2.5.3). The Area under the curve of the receiver operating characteristic curve (AUC ROC) was estimated using the scikit-learn scientific computing library (version 0.22.1), based on the Mann-Whitney U-statistic. The 95\% confidence intervals of the true AUC was estimated using DeLong's method implemented by pROC (version 1.16.2) in R. \\
\noindent\textbf{WSI Processing} \\
\textit{Segmentation.} Tissue segmentation of WSIs was performed automatically using the CLAM library at a downsampled magnification of each slide. A binary mask for the tissue regions were computed by applying binary thresholding to the saturation channel of the image downsample after conversion from RGB to the HSV color space. Median blurring and morphological closing were also performed to smooth the detected tissue contours and suppress artifacts such as small gaps and holes. The approximate contours of the detected tissue as well as tissue cavities were then filtered based on their area to produce the final segmentation mask. \\
\textit{Patching.}
We exhaustively cropped segmented tissue contours into 256 $\times$ 256 patches (without overlap) at 20$\times$ magnification (if the 20$\times$ downsample is not found in the image pyramid, 512 $\times$ 512 patches were instead cropped from the 40$\times$ downsample and downscaled to 256 $\times$ 256), leading to a total of ~280 million cropped patches (median: 10320 patches per slide). We refer to the collective set of all patches (known as instances) extracted from a particular WSI as a bag. \\
\textit{Feature Extraction.}
Given the enormous bag sizes (number of patches in each WSI) in our dataset,
we first used a convolutional neural network based on the ResNet50 architecture to encode each patch into a compact low-dimensional feature vector. Specifically, a ResNet50 model pretrained on Imagenet was truncated after the 3rd residual block and was followed by an adaptive mean-spatial pooling layer to reduce the spatial feature map obtained from each 256 $\times$ 256 $\times$ 3 RGB image patch into a descriptive, one-dimensional feature representation of length of 1024. To perform this feature extraction step efficiently, we used up to 16 GPUs in parallel with a batch-size of 128 per GPU. \\
\noindent\textbf{Interpreting Model Prediction via Attention Heatmap} \\
To visually interpret the importance of each region in a WSI towards the model's classification predictions, we first computed the reference distribution of attention scores by tiling the WSI into 256 $\times$ 256 patches without overlap and computing the attention score for each patch for the task of primary origin prediction. To generate more fine-grained heatmaps, we subsequently repeated the tiling but with an overlap of up to 90\% and converted the attention scores computed from overlapping crops to normalized percentile scores between 0.0 (low attention) to 1.0 (high attention) based on the initial reference distribution. The normalized scores were then registered onto the original WSI corresponding each patch's spatial location and scores in overlapped regions were accumulated and averaged. Finally, a colormap was applied to the attention scores and the heatmap was displayed as an overlay layer with a transparency value of 0.5. These attention maps have been shown in \textbf{Figure 3, Extended Data Figure 6, 8, 9} and can also be visulized in our interactive demo http://toad.mahmoodlab.org.

\vspace{-6mm}
\section*{Data Availability}
\vspace{-6mm}
Digitized, high-resolution diagnostic whole slide image data from the TCGA and their corresponding diagnoses are publicly accessible through the NIH genomic data commons. All reasonable requests for in-house raw and analyzed data and materials will be promptly reviewed by the authors to determine whether the request is subject to
any intellectual property or confidentiality obligations. Patient-related data not included in the paper may be subject to patient confidentiality. All requests for data that can be shared will be processed through formal channels, in concordance with institutional and departmental guidelines and will require a material transfer agreement. 
\vspace{-4mm}
\section*{Code Availability}
\vspace{-6mm}
All code was implemented in Python using PyTorch as the primary deep learning package. All code and scripts to reproduce the experiments of this paper are available at \url{https://github.com/mahmoodlab/TOAD}
All source code is provided under the GNU GPLv3 free software license.
\vspace{-6mm}

\section*{Author Contributions}
\vspace{-4mm}
M.Y.L. and F.M. conceived the study and designed the experiments. M.Y.L. performed the experimental analysis. M.Z. D.W. T.C. curated the in-house datasets. M.Y.L. M.Z. M.S. J.L. F.M. analyzed the results. M.Y.L M.S. J.L. developed data visualization tools. M.Y.L. F.M. prepared the manuscript. F.M. supervised the research.
\vspace{-4mm}
\section*{Acknowledgements}
\vspace{-6mm}
The authors would like to thank Alexander Bruce for scanning internal cohorts of patient histology slides at BWH; Jingwen Wang, Matteo Barbieri, Katerina Bronstein, Lia Cirelli, Eric Askeland for querying the BWH slide database and retrieving archival slides; Celina Li for assistance with EMRs and RPDR; Martina Bragg, Terri Mellen and Sarah Zimmet for logistical support; Zahra Noor for developing the interactive demo website; and Kai-ou Tung of Boston Children's Hospital for anatomical illustrations. This work was supported in part by internal funds from BWH Pathology, Google Cloud Research Grant and Nvidia GPU Grant Program and NIGMS R35GM138216 (F.M.). M.S. was additionally supported by the NIH Biomedical Informatics and Data Science Research Training Program, grant number: NLM T15LM007092. The content is solely the responsibility of the authors and does not reflect the official views of the National Institute of Health, National Institute of General Medical Sciences or the National Library of Medicine.
\vspace{-6mm}
\section*{Competing Interests}
\vspace{-6mm}
The authors declare that they have no competing financial interests.
\vspace{-6mm}

\section*{Ethics Oversight}
\vspace{-6mm}
The study was approved by the Mass General Brigham (MGB) IRB office under protocol 2020P000233.
\end{spacing}

\begin{nolinenumbers}
\vspace{-9mm}

\section*{References} 
\vspace{2mm}

\begin{spacing}{0.9}
\bibliographystyle{naturemag}
\bibliography{sample}
\end{spacing}
\end{nolinenumbers}

\begin{figure*}
\vspace{-9mm}
\includegraphics[width=\textwidth]{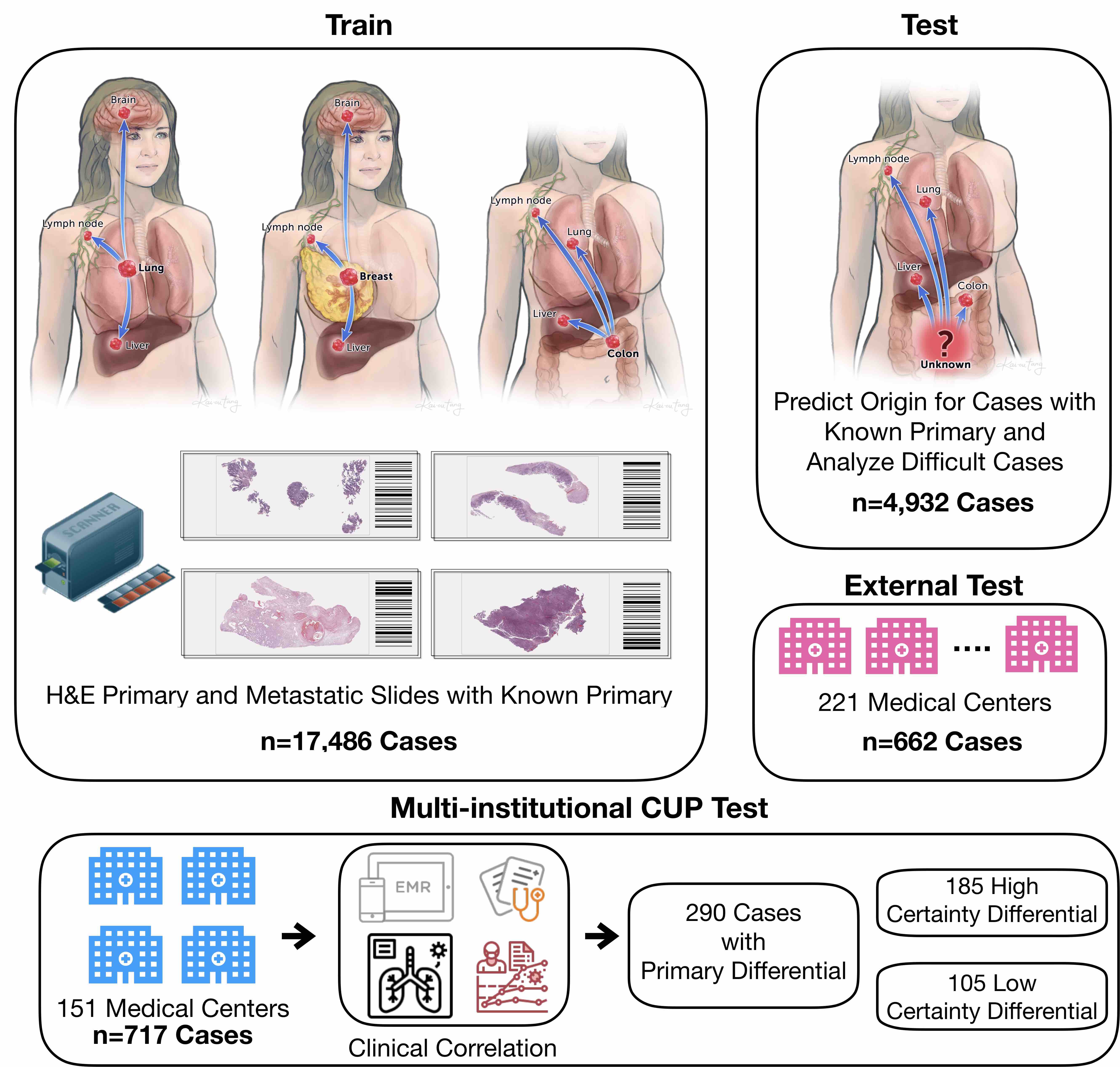}
\caption*{\textbf{Extended Data Figure 1. Overall study design.} \textbf{1. Model training and testing.} For model development, we collected in total, 24,885 FFPE H\&E digitized diagnostic slides (from 23,297 patient cases) of confirmed diagnosis and randomly sampled 70\% of cases (17,486 slides) for training the model, and 20\% of cases (4,932 slides) as a held-out for evaluation. The remaining 10\% of cases (2,467 slides) were used for validation during training in order the select the best performing model. \textbf{2. External test.} In order to further assess the model's ability to generalize on data from sources not encountered during training, we also evaluated the model on an external test cohort of 662 cases, submitted for consultation from over 200 US and international medical centers. \textbf{3. Evaluation on challenging CUP cases.} Lastly, to assess the model's ability to inform meaningful predictions for origins of cancers that cannot be readily diagnosed by human experts, we curated an additional diverse dataset of 717 CUP cases sourced from institutions across the country and outside the US. While the primary could not be initially assigned for all of these cases based on H\&E histology alone, using EMR and evidence from many other forms of clinical and diagnostic reports, we identified a subset of 290 cases for which a primary differential was eventually assigned over the course of the patient's history. We validated our model against the recorded primary differential for agreement, showcasing its applicability to cases without clear morphological indication for a particular primary.}
\end{figure*}

\begin{figure*}
\includegraphics[width=\textwidth]{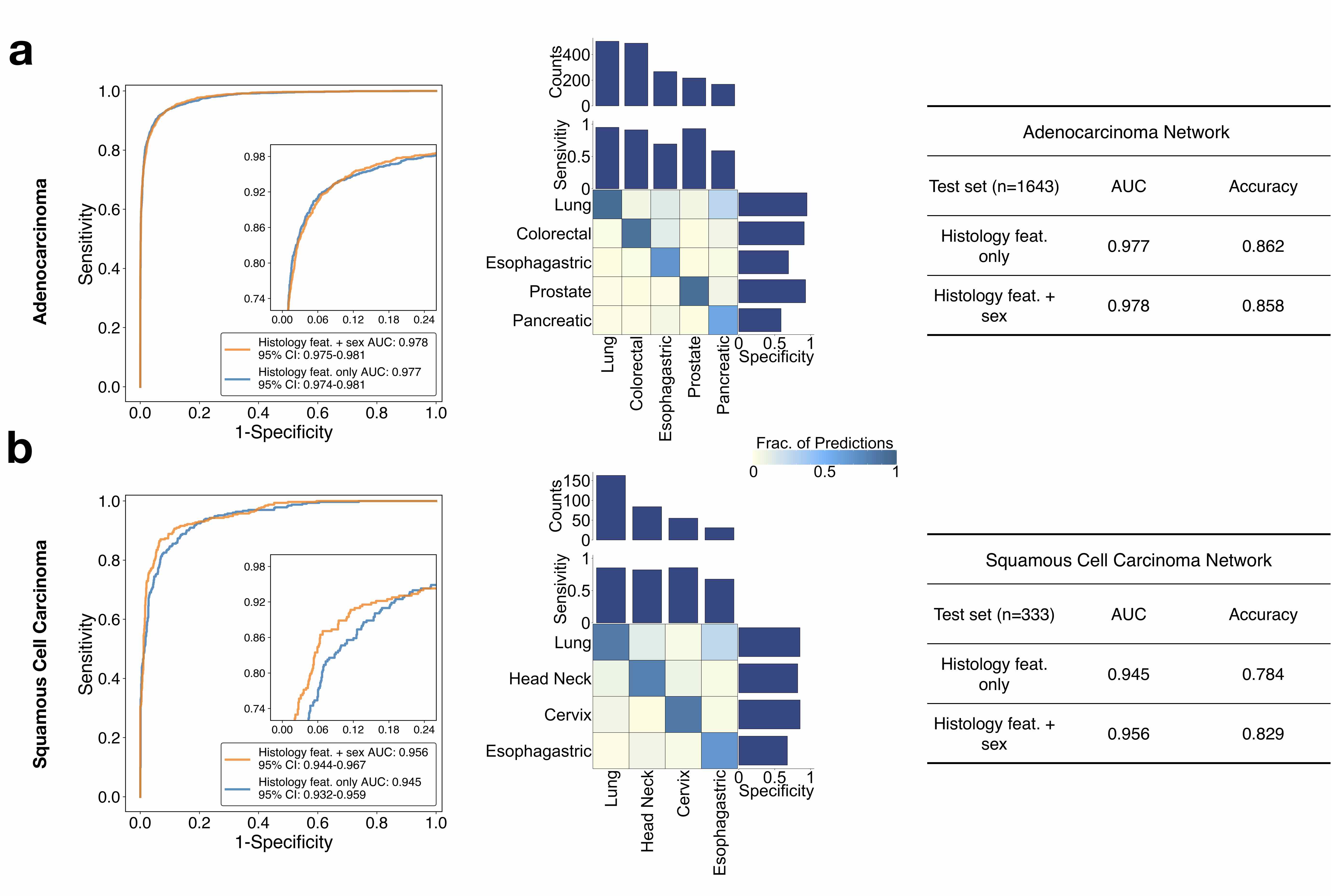}
\caption*{\textbf{Extended Data Figure 2. Classification performance of adenocarcinoma network and squamous cell carcinoma network.} Often pathologists can readily distinguish between adenocarcinoma (AD) and squamous cell carcinoma (SCC) based on the morphological and architectural appearance of the tumor cells present in the tissue. However, within the respective family of AD and SCC subtypes, determining the origin of the tumor can remain a challenging task. Therefore we hypothesized that we can develop TOAD models to specifically predict the origin of tumors for top primary sites of adenocarcinoma \textbf{a.} and SCC \textbf{b.}. Cases from five and four primary sites were chosen for the development of the AD classifier and SCC classifier respectively, based on their frequency in the database. The confusion matrix is plotted for each TOAD model (\textbf{middle}). Additionally, micro-averaged AUC and overall accuracy are noted for the models trained with and without incorporating sex as an additional covariate (\textbf{left, right}). The AD network achieved an micro-averaged AUC ROC of 0.977 (95\% CI: 0.974 - 0.981) and overall accuracy of 85.8\% and did not benefit from adding sex, where the model achieved a similar AUC ROC of 0.978 (95\% CI: 0.975 - 0.981) and accuracy of 86.2\%. The SCC network scored a higher sensitivity for cervical cancer (0.85 with sex vs. 0.69 without sex), which led to a modest increase in AUC from 0.945 (95\% CI: 0.932 - 0.959) to 0.956 (95\% CI: 0.944 - 0.967) and accuracy of 78.4\% to 82.9\%. }
\end{figure*}

\begin{figure*}
\includegraphics[width=\textwidth]{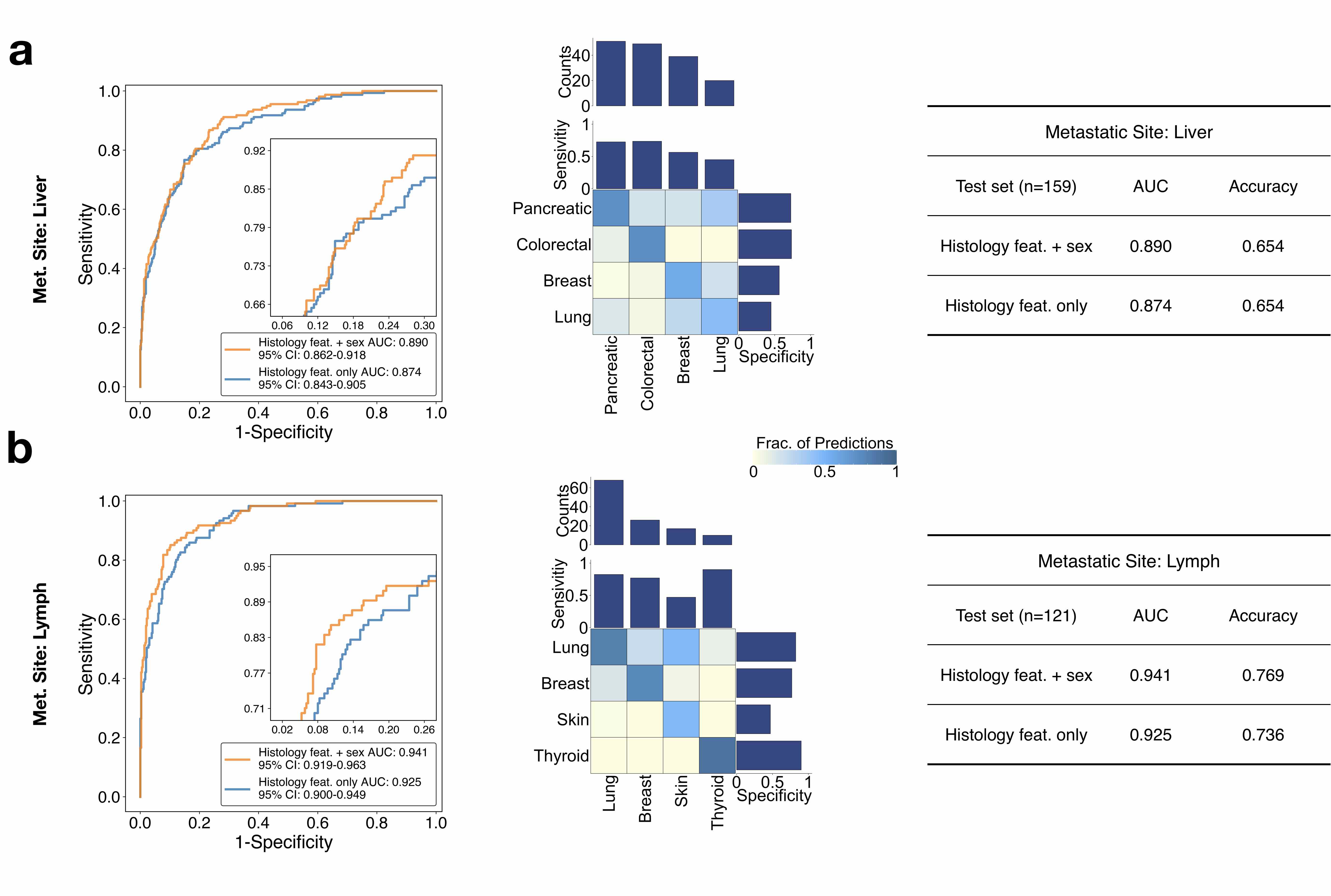}
\caption*{\textbf{Extended Data Figure 3. Classification performance of site-specific networks for the tumor metastasized to the liver and lymph node.} We also explored the possibility of using TOAD to predict the primary origins of metastatic tumors grouped by a common metastatic site, including the liver (\textbf{a.}) and the lymph node (\textbf{b.}).
Metastatic cases from the top four primary origins for each site were chosen based on their frequency in our database. \textbf{left.} Micro-averaged ROC curve. \textbf{middle.} Confusion matrix. \textbf{right.} Overall accuracy and micro-averaged AUC. For tumors metastasized to the liver, the micro-averaged AUC ROC was 0.890 (95\% CI: 0.862 - 0.918) when incorporating sex vs. 0.874 (95\% CI: 0.843 - 0.905) without sex as an additional covariate. We found that while incorporating sex improved the sensitivity for breast cancer (0.87 with sex vs. 0.56 without sex), it came at the expense of lowered sensitivity for all other primary sites. On the other hand, incorporating sex led to a substantial increase in the sensitivities for lung and breast cancers metastasized to the lymph node and the overall accuracy of the lymph node network increased from 73.6\% to 76.9\%.}
\end{figure*}

\clearpage
\begin{figure*}
\includegraphics[width=\textwidth]{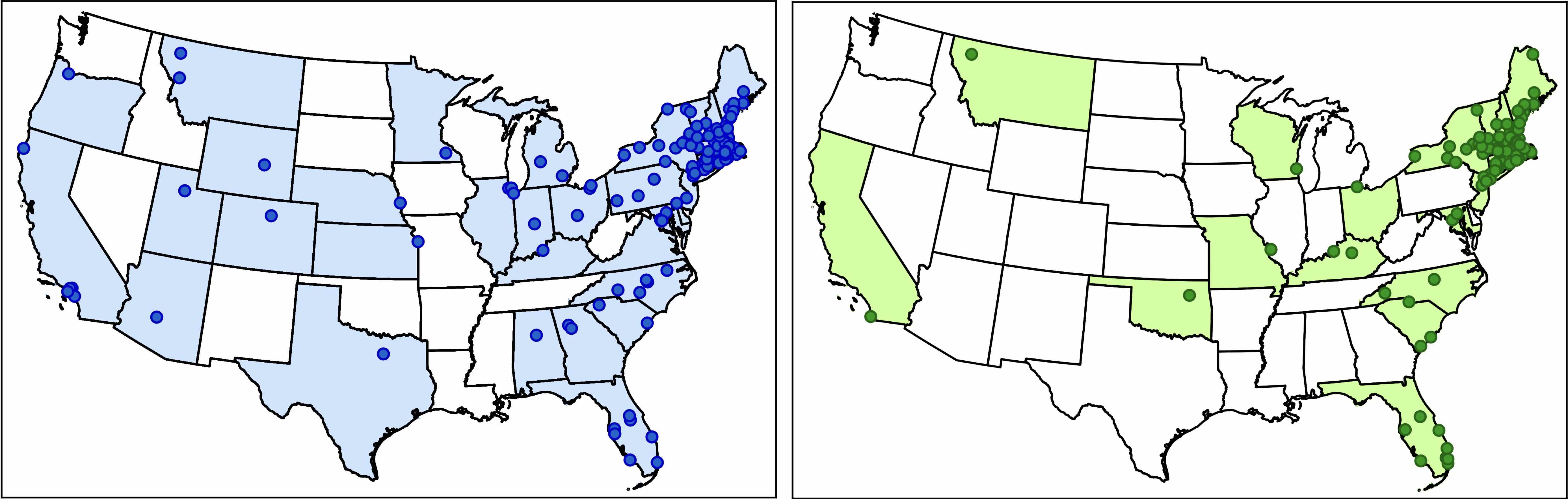}
\caption*{\textbf{Extended Data Figure 4. Geographical diversity of our external test set and CUP cases.} \textbf{left.} The external test set of 662 cases are submitted from in total 202 medical centers across 34 states in USA and 19 medical centers from 8 other countries including Switzerland, Brazil, Greece, United Arab Emirates, China, Saudi Arabia, Kuwait and Canada. \textbf{right.} Similarly, our CUP cases consist of 717 slides from 146 medical centers across 22 US states and 5 centers from 2 other countries including China and Kuwait.}
\end{figure*}

\clearpage
\begin{figure*}
\includegraphics[width=\textwidth]{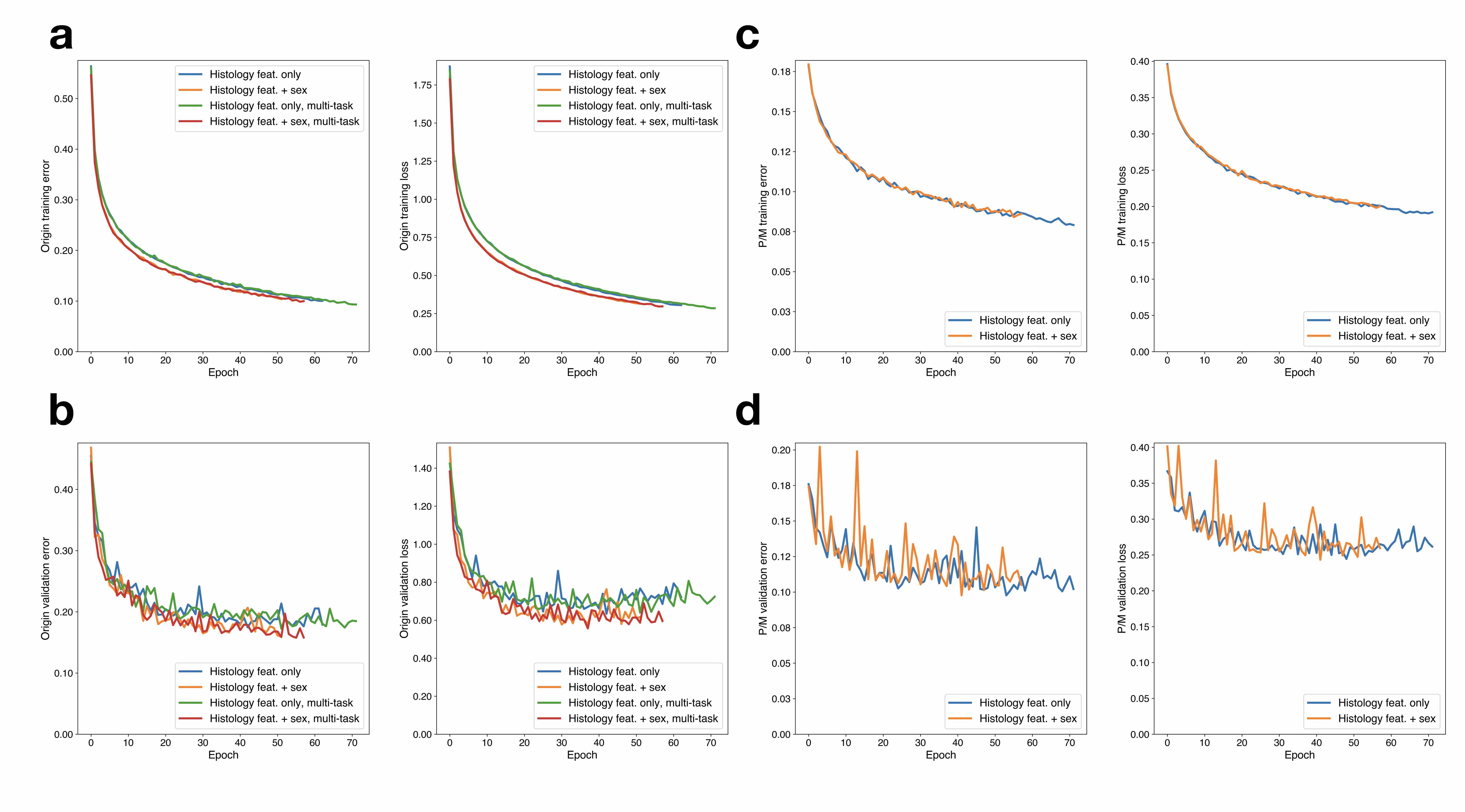}
\vspace{-8mm}
\caption*{\textbf{Extended Data Figure 5. Model performance during training and validation.} \textbf{a, b.} Classification error and cross-entropy loss for predicting the tumor origin for different model configurations (averaged over each epoch) on the training and validation set respectively. \textbf{c, d.} Classification error and cross-entropy loss for predicting primary vs. metastatic tumor for multi-task (and single task) model configurations.}
\end{figure*}
\clearpage

\begin{figure*}
\includegraphics[width=\textwidth]{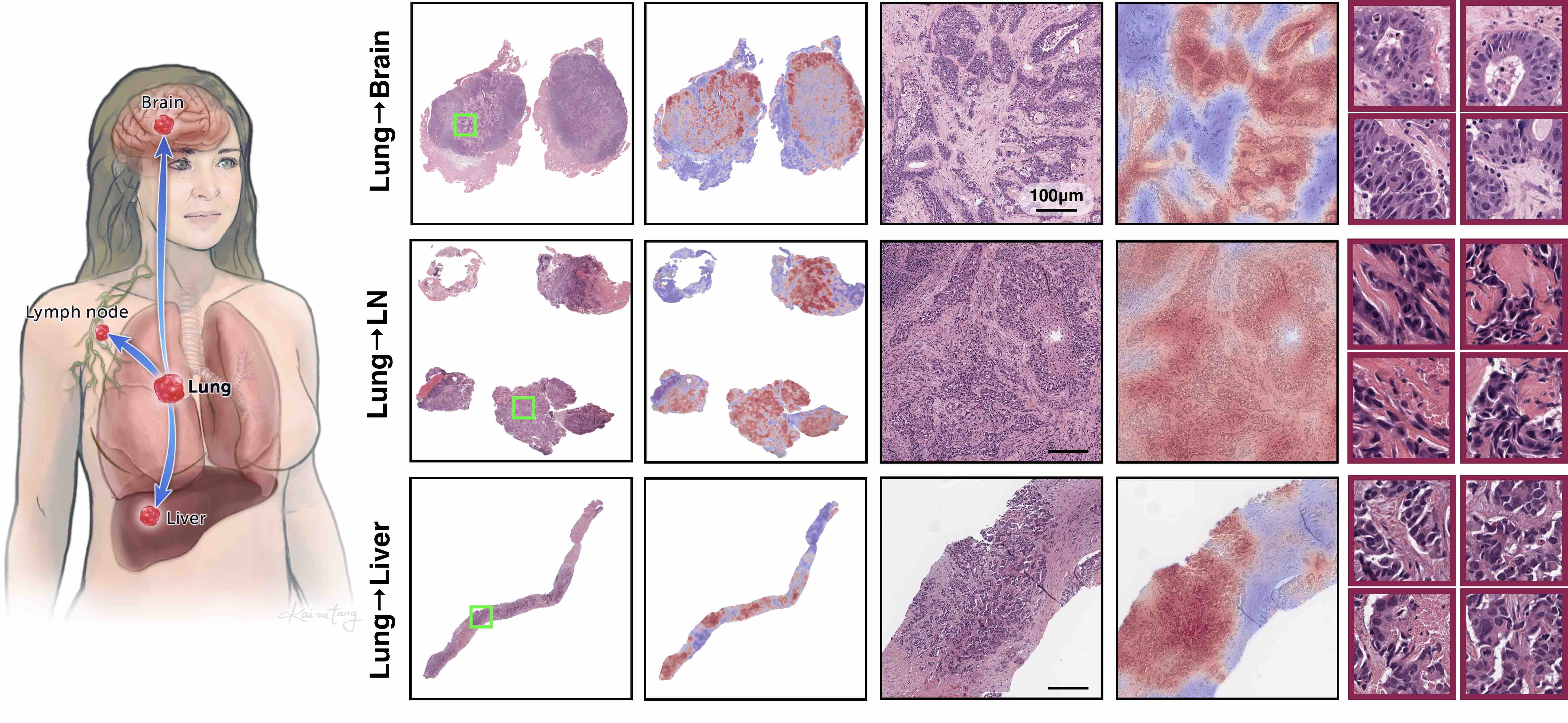}

\caption*{\textbf{Extended Data Figure 6. Exemplars of metastases from lung primaries with attention heatmaps}. From left to right, low magnification with corresponding attention map, medium magnification with corresponding attention map, and high magnification patches. Medium and high magnification views demonstrate sheets of cells, variably-sized glands, and cells in infiltrative single files. The cells have large, hyperchromatic nuclei and low nuclear:cytoplasmic ratio, consistent with metastatic lung carcinomas.}
\end{figure*}
\clearpage

\begin{figure*}
\includegraphics[width=\textwidth]{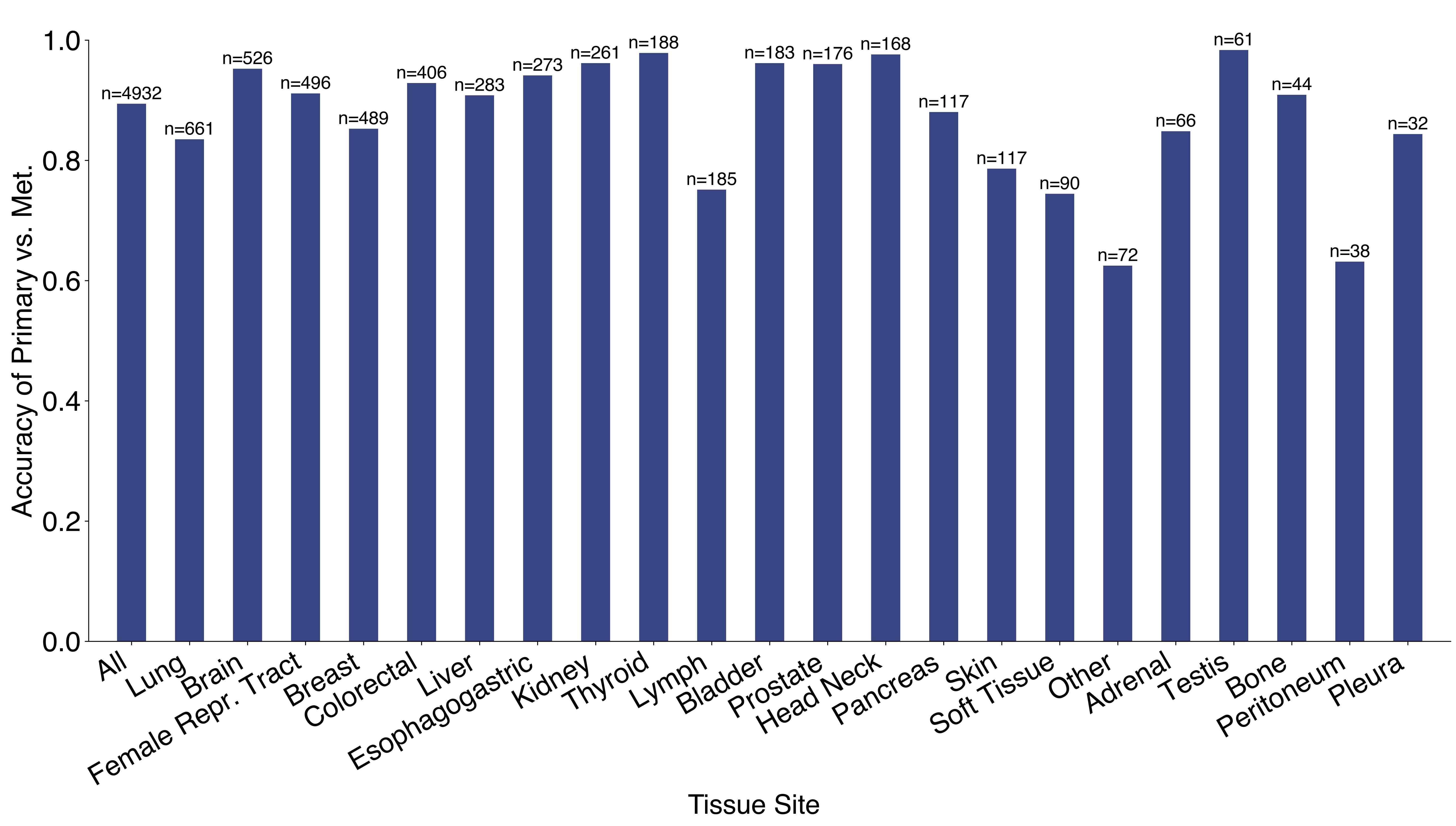}

\caption*{\textbf{Extended Data Figure 7. Model performance on the binary problem of distinguishing between primary and metastatic tumors in different tissue sites.} The barplot shows model accuracy (y-axis) on the test set (n=4932) for different tissue sites (x-axis) and the number of cases found at each site. These sites should not be confused with the 18 common primary sites used for the origin determination task. This bar plot was plotted by stratifying all test cases based on the site from which the tissue was sampled and the accuracy reported is for predicting if the slide is a primary or metastatic tumor.}
\end{figure*}
\clearpage

\begin{figure*}
\includegraphics[width=\textwidth]{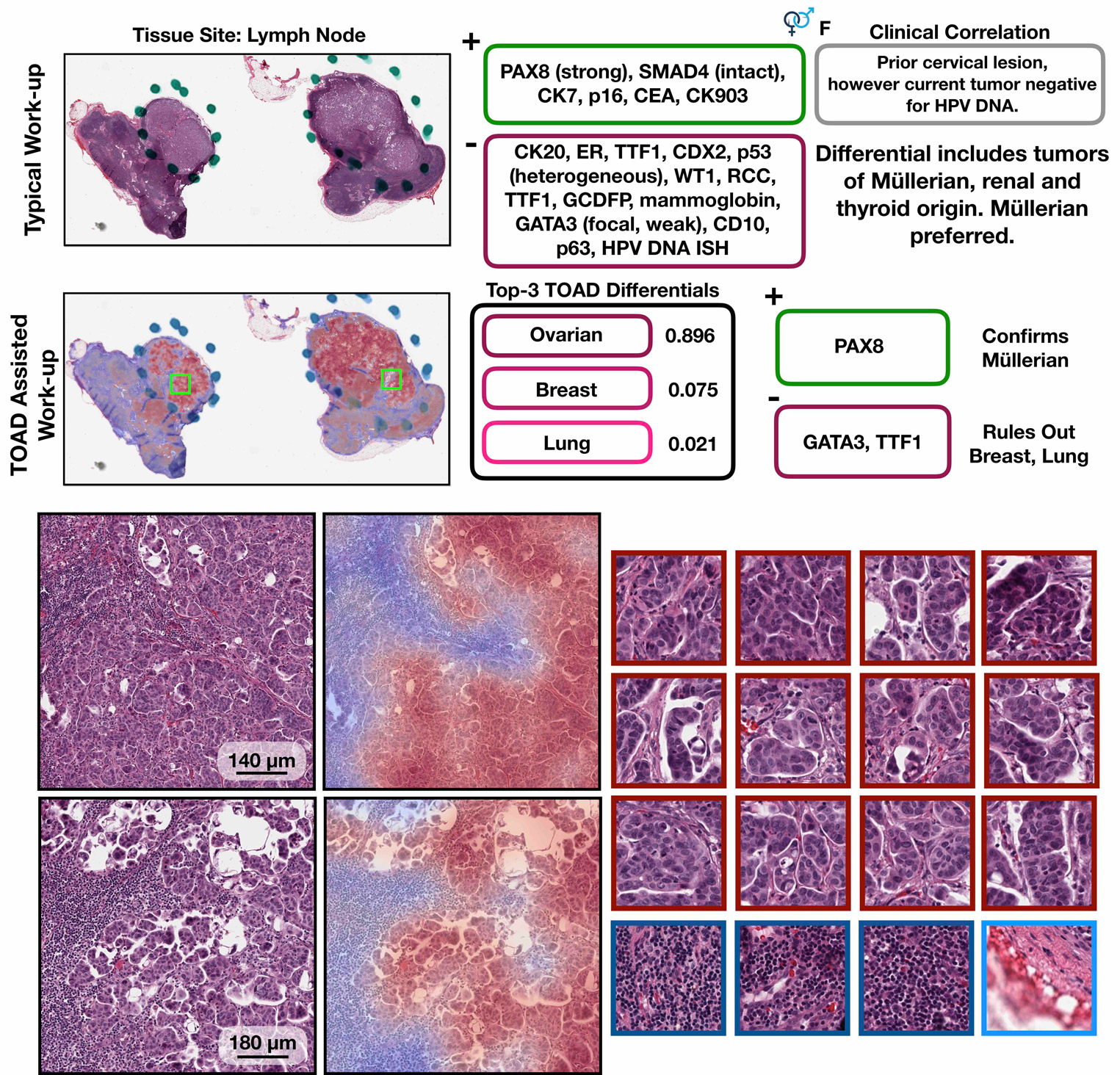}

\caption*{\textbf{Extended Data Figure 8. TOAD-assisted CUP work-up: case study 1.} The figure above shows a representative case which underwent a standard CUP work-up involving extensive IHC staining and clinical correlation. Strong PAX8 staining suggested Müllerian origin and multiple IHCs were used to rule out other primaries. Retrospectively, we analyzed the case with TOAD and found the top-3 determinations to be Ovarian, Breast, Lung, and following this deterimation just three IHC stains (PAX8, GATA3, and TTF1) could be used to confirm Müllerian origin and rule out breast carcinoma and lung adenocarcinom respectively. This workflow demonstrates how TOAD can be used as an assistive diagnostic tool. 
.}
\end{figure*}
\clearpage

\begin{figure*}
\includegraphics[width=\textwidth]{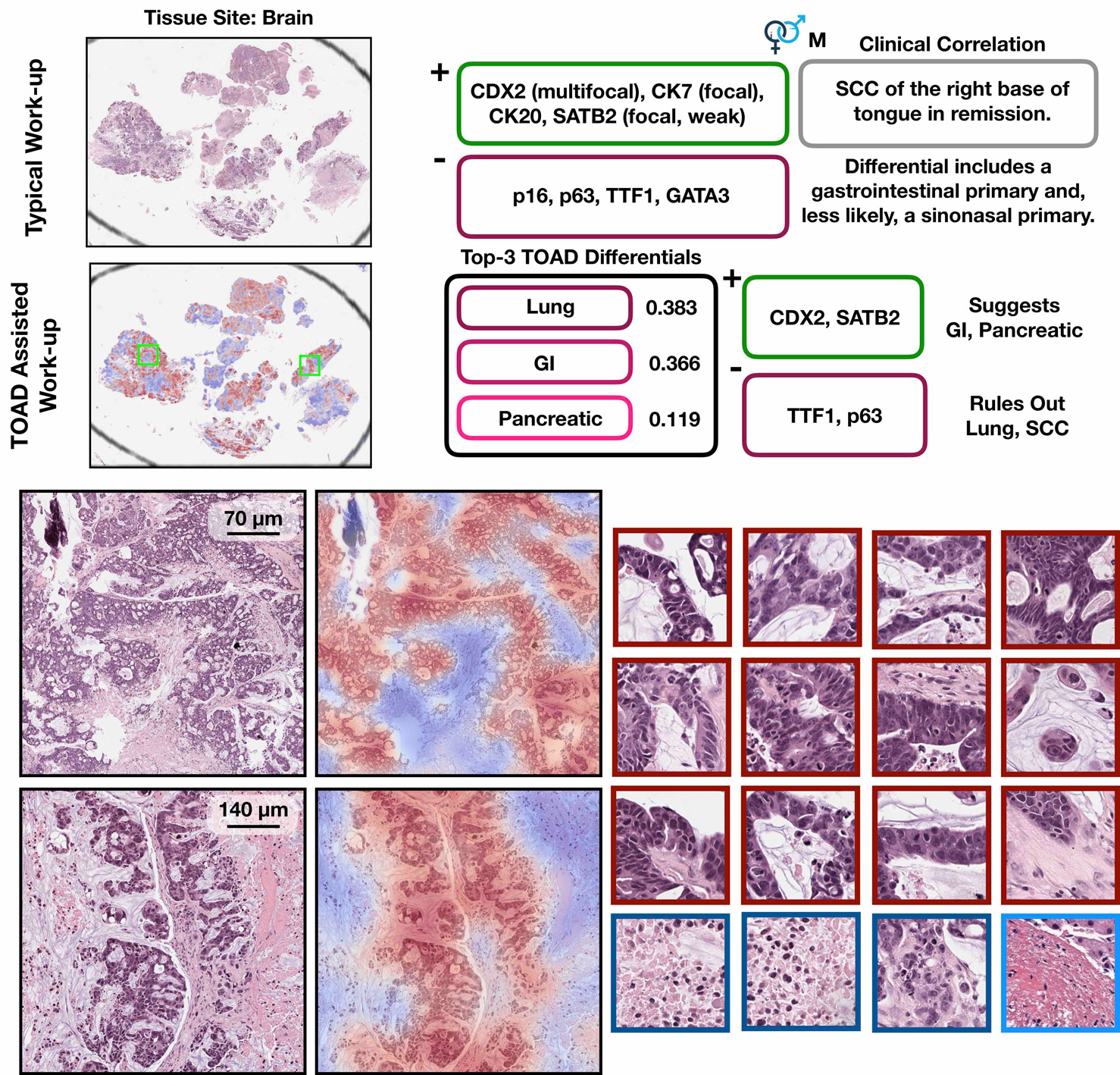}

\caption*{\textbf{Extended Data Figure 9. TOAD-assisted CUP work-up: case study 2.} This representative case demonstrates that TOAD can be used as an assistive tool or as an additional reader even when the top-1 prediction is not in agreement with the differential assigned. This particular case of brain metastasis underwent a typical CUP work-up with several stains and clinical correlation. Retrospectively, when we analyzed the case using CUP we found that the top-3 predictions included Lung, GI, Pancreatic in decreasing order of confidence, where the confidence between lung and GI was almost the same. TTF1 could be used to rule out Lung, p63 could be used to rule out SCC because of prior SCC history and optionally CDX2 and SATB2 could be used to confirm GI origin. Additionally since adenocarcinoma morphology is identified, we tested this case using our adenocarcinoma TOAD model (see \textbf{Extended Data Figure 2}). This model suggested similar prediction scores for Lung ($p=0.351$), Esophagogastric ($p=0.345$), but listed Colorectal ($p=0.269$) as the other likely candidate, whereas Pancreatic did not appear within the top-3 predicted origins, accompanied by a probability score of only $p=0.034$. This suggests it may be possible to also consider predictions from more specific networks (\textit{e.g.} metastatic-site-specific, or morphological-subtype-specific) when trying to rule out plausible candidates.}
\end{figure*}
\clearpage

\begin{table}
\caption*{\textbf{Extended Data Table 1. WSI Dataset summary}}
\centering
\begin{tabular}{lp{3cm}p{3cm}p{2.5cm}p{3cm}l}
\toprule
Primary   Organ & Training & Validation & Test & External Test & Total \\
\midrule
Lung            & 2800     & 405        & 792  & 83            & 4080  \\
Breast          & 2286     & 328        & 651  & 70            & 3335  \\
Colorectal      & 1718     & 244        & 486  & 212           & 2660  \\
Ovarian         & 778      & 109        & 220  & 20            & 1127  \\
Pancreatic      & 628      & 88         & 182  & 11            & 909   \\
Adrenal         & 198      & 21         & 55   & 0             & 274   \\
Skin            & 693      & 97         & 189  & 52            & 1031  \\
Prostate        & 774      & 120        & 212  & 20            & 1126  \\
Renal           & 983      & 136        & 284  & 11            & 1414  \\
Bladder         & 758      & 102        & 210  & 22            & 1092  \\
Esophagogastric   & 1046     & 148        & 298  & 20            & 1512  \\
Thyroid         & 708      & 100        & 202  & 2             & 1012  \\
Head Neck       & 626      & 86         & 178  & 10            & 900   \\
Glioma          & 1649     & 235        & 446  & 111           & 2441  \\
Germ Cell       & 256      & 28         & 66   & 3             & 353   \\
Endometrial     & 1009     & 140        & 297  & 11            & 1457  \\
Cervix          & 262      & 34         & 75   & 2             & 373   \\
Liver           & 314      & 46         & 89   & 2             & 451   \\
\midrule
\textbf{Total}           & \textbf{17486}    & \textbf{2467}       & \textbf{4932} & \textbf{662}           & \textbf{25547} \\
\bottomrule
\end{tabular}
\end{table}

\clearpage

\begin{longtable}{p{3cm}p{14cm}}
\caption*{\textbf{Extended Data Table 2. Tumor types grouped for classification}}\\
\toprule
 Primary Organ &                                         Disease Models Included \\
\midrule
\endhead
\endfoot

\bottomrule
\endlastfoot
          Lung &                                                         Lung Adenocarcinoma, Lung Squamous Cell Carcinoma, Non-Small Cell Lung Cancer, Small Cell Lung Cancer, Poorly Differentiated Non-Small Cell Lung Cancer, Lung Carcinoid, Large Cell Neuroendocrine Carcinoma, Atypical Lung Carcinoid, Lung Adenosquamous Carcinoma, Lung Neuroendocrine Tumor, Sarcomatoid Carcinoma of the Lung, Large Cell Lung Carcinoma  \\ \midrule
        Breast &                                                                                                                                                                                                  Breast Invasive Ductal Carcinoma, Invasive Breast Carcinoma, Breast Invasive Lobular Carcinoma, Breast Mixed Ductal and Lobular Carcinoma, Breast Invasive Mixed Mucinous Carcinoma, Breast Ductal Carcinoma In Situ \\ \midrule
    Colorectal &                                                                                                                                                                                                                                                                                               Colon Adenocarcinoma, Rectal Adenocarcinoma, Colorectal Adenocarcinoma, Mucinous Adenocarcinoma of the Colon and Rectum \\ \midrule
        Glioma &                                                                                                                                                                                    Glioblastoma, Glioblastoma Multiforme, Astrocytoma, Diffuse Glioma, Anaplastic Astrocytoma, Oligodendroglioma, Pilocytic Astrocytoma, Anaplastic Oligodendroglioma, Ganglioglioma,  Anaplastic Oligoastrocytoma, Oligoastrocytoma  \\ \midrule
 Esophagogastric &                                                                                                                                                          Esophageal Adenocarcinoma, Stomach Adenocarcinoma, Esophagogastric Adenocarcinoma, Adenocarcinoma of the Gastroesophageal Junction, Esophageal Squamous Cell Carcinoma, Diffuse Type Stomach Adenocarcinoma, Poorly Differentiated Carcinoma of the Stomach  \\ \midrule
   Endometrial &                                                                                                                            Uterine Endometrioid Carcinoma, Uterine Serous Carcinoma/Uterine Papillary Serous Carcinoma, Endometrial Carcinoma, Uterine Carcinosarcoma/Uterine Malignant Mixed Mullerian Tumor, Uterine Mixed Endometrial Carcinoma, Uterine Clear Cell Carcinoma, Uterine Undifferentiated Carcinoma  \\ \midrule
         Renal &                                                                                                                                                                        Renal Clear Cell Carcinoma, Renal Cell Carcinoma, Papillary Renal Cell Carcinoma, Chromophobe Renal Cell Carcinoma, Renal Oncocytoma, Collecting Duct Renal Cell Carcinoma, Renal Non-Clear Cell Carcinoma, Unclassified Renal Cell Carcinoma  \\ \midrule
       Ovarian &  High-Grade Serous Ovarian Cancer, Endometrioid Ovarian Cancer, Clear Cell Ovarian Cancer, Low-Grade Serous Ovarian Cancer, Serous Ovarian Cancer, Ovarian Epithelial Tumor, Ovarian Carcinosarcoma/Malignant Mixed Mesodermal Tumor, Mucinous Ovarian Cancer, Serous Borderline Ovarian Tumor, Ovarian Cancer, Other, Mixed Ovarian Carcinoma, Mucinous Borderline Ovarian Tumor, Small Cell Carcinoma of the Ovary  \\ \midrule
      Prostate &                                                                                                                                                                                                                                                                                                                                                                Prostate Adenocarcinoma, Prostate Small Cell Carcinoma \\ \midrule
       Bladder &                                                                                                                                                                                                                                                                                               Bladder Urothelial Carcinoma, Upper Tract Urothelial Carcinoma. Bladder Adenocarcinoma, Bladder Squamous Cell Carcinoma \\ \midrule
       Thyroid &                                                                                                                                                                                                                                                                      Papillary Thyroid Cancer, Medullary Thyroid Cancer, Anaplastic Thyroid Cancer, Hurthle Cell Thyroid Cancer, Poorly Differentiated Thyroid Cancer \\ \midrule
      Skin &                                                                                                                                                                                                                                                                                                                                                                                         Melanoma, Cutaneous Melanoma  \\ \midrule
    Pancreatic &                                                                                                                                                                                                                                                                    Pancreatic Adenocarcinoma, Adenosquamous Carcinoma of the Pancreas, Intraductal Papillary Mucinous Neoplasm, Acinar Cell Carcinoma of the Pancreas \\ \midrule
 Head and Neck &                                                                                                                                                                                                                                    Oral Cavity Squamous Cell Carcinoma, Oropharynx Squamous Cell Carcinoma, Head and Neck Squamous Cell Carcinoma, Larynx Squamous Cell Carcinoma, Sinonasal Squamous Cell Carcinoma  \\ \midrule
         Liver &                                                                                                                                                                                                                                                                                                                                                                                              Hepatocellular Carcinoma \\ \midrule
        Cervix &                                                                                                                                                                                                                                                                                              Cervical Squamous Cell Carcinoma, Endocervical Adenocarcinoma, Cervical Adenocarcinoma, Cervical Adenosquamous Carcinoma \\ \midrule
        Germ Cell & 
        Seminoma, Mixed Germ Cell Tumor,  Yolk Sac Tumor, Embryonal Carcinoma, Teratoma, Mature Teratoma,  Non-Seminomatous Germ Cell Tumor \\ \midrule
       Adrenal &                                                                                                                                                                                                                                                                                                                                                                      Adrenocortical Carcinoma, Adrenocortical Adenoma \\
\end{longtable}

\clearpage

\begin{longtable}{lp{2cm}p{2cm}p{2cm}p{5cm}p{2cm}}
\caption*{\textbf{Extended Data Table 3. Test performance on 18-class classification of primary origin}}\\
\toprule
Primary Origin &  Precision &  Recall &  F1-score &  AUC-ROC (95\% CI) & Count \\
\midrule
\endhead
\midrule
\midrule
\endfoot
\bottomrule
\endlastfoot
Lung          &     0.778 &  0.808 &    0.793 &  0.970  \enspace (0.965 - 0.975) &      792 \\
Breast        &     0.879 &  0.873 &    0.876 &  0.988  \enspace (0.985 - 0.992) &      651 \\
Colorectal    &     0.928 &  0.877 &    0.902 &  0.991  \enspace (0.987 - 0.995) &      486 \\
Glioma        &     0.975 &  0.951 &    0.963 &  0.999  \enspace (0.998 - 1.000) &      446 \\
Esophagogastric &     0.819 &  0.715 &    0.763 &  0.964  \enspace (0.952 - 0.975) &      298 \\
Endometrial   &     0.895 &  0.778 &    0.832 &  0.986  \enspace (0.979 - 0.992) &      297 \\
Renal         &     0.892 &  0.898 &    0.895 &  0.989  \enspace (0.981 - 0.996) &      284 \\
Ovarian       &     0.651 &  0.805 &    0.720 &  0.980  \enspace (0.974 - 0.986) &      220 \\
Prostate      &     0.763 &  0.925 &    0.836 &  0.992  \enspace (0.987 - 0.997) &      212 \\
Bladder       &     0.843 &  0.743 &    0.790 &  0.983  \enspace (0.975 - 0.991) &      210 \\
Thyroid       &     0.902 &  0.911 &    0.906 &  0.995  \enspace (0.991 - 0.998) &      202 \\
Skin          &     0.791 &  0.783 &    0.787 &  0.984  \enspace (0.977 - 0.990) &      189 \\
Pancreatic    &     0.605 &  0.808 &    0.692 &  0.971  \enspace (0.958 - 0.983) &      182 \\
Head Neck     &     0.897 &  0.781 &    0.835 &  0.988  \enspace (0.982 - 0.995) &      178 \\
Liver         &     0.806 &  0.843 &    0.824 &  0.996  \enspace (0.993 - 1.000) &       89 \\
Cervix        &     0.759 &  0.587 &    0.662 &  0.978  \enspace (0.963 - 0.993) &       75 \\
Germ Cell     &     0.866 &  0.879 &    0.872 &  0.997  \enspace (0.994 - 0.999) &       66 \\
Adrenal       &     0.909 &  0.727 &    0.808 &  0.996  \enspace (0.992 - 1.000) &       55 \\
\midrule
Micro-avg       &     0.836 &  0.836 &    0.836 &  0.988  \enspace (0.987 - 0.990) &     4932 \\
Macro-avg     &     0.831 &  0.816 &    0.820 &                    0.986 &     4932 \\
Weighted-avg  &     0.843 &  0.836 &    0.837 &                    0.984 &     4932 \\
\end{longtable}

\newpage

\begin{table}[H]
\centering
\caption*{\textbf{Extended Data Table 4. Ablation study}}
\subcaption*{\textbf{A. Performance on Origin Prediction}}
\begin{tabular}{lp{6cm}lllp{2.75cm}l}
\toprule
Test set (n=4932)             & Micro-Avg AUC (95\% CI)   & Top-1 Acc & Top-3 Acc & Top-5 Acc  \\ \midrule
Hist. feat. + sex, multi-task & 0.988 (0.987 - 0.989) & 0.836     & 0.944     & 0.978        \\
Hist. feat. only,  multi-task & 0.986 (0.984 - 0.987) & 0.828     & 0.939     & 0.968       \\
Hist. feat. + sex             & 0.988 (0.987 - 0.989) & 0.825     & 0.945     & 0.976         \\
Hist. feat. only              & 0.987 (0.985 - 0.988)& 0.824     & 0.939     & 0.971          \\
\bottomrule
\end{tabular}
\bigskip

\subcaption*{\textbf{B. Performance on Primary vs. Metastatic Prediction}}
\begin{tabular}{lp{4cm}lllp{2.75cm}l}
\toprule
Test set (n=4932)  & AUC (95\% CI)   & Accuracy \\ \midrule
Hist. feat. + sex, multi-task & 0.934 (0.926 - 0.942) & 0.894    \\
Hist. feat. only,  multi-task & 0.930 (0.922 - 0.938) & 0.884    \\
\bottomrule
\end{tabular}
\end{table}
\end{document}